\documentclass[10pt,twocolumn,twoside]{IEEEtran}
  \IEEEoverridecommandlockouts                              % This command is only
  \usepackage{graphicx}
  \usepackage{psfrag}
  \usepackage{lipsum}
  \usepackage[utf8]{inputenc}
  \usepackage{commath}
  \usepackage{amsfonts}
  \usepackage{amsthm}
  \usepackage{xcolor}
  \usepackage{xpatch}
  \usepackage{capt-of}
\xapptocmd{\appendix}{%
}{}{\PatchFailed}
  \newtheorem{theorem}{Theorem}[section]
  \newtheorem{corollary}[theorem]{Corollary}
  \newtheorem{lemma}[theorem]{Lemma}
  \newtheorem{proposition}[theorem]{Proposition}
  \newtheorem{remark}{Remark}

 \usepackage[noadjust]{cite}
  \usepackage{breqn}
  \usepackage{mathtools}
  \usepackage{hhline}
  \usepackage{threeparttable}

  \usepackage{amsmath}
  
  \newcommand{\argmax}{\arg\max}
  
  \newcommand\taua{\underline{\tau}^{\mathrm{A}}}
  \newcommand\taud{\underline{\tau}^{\mathrm{D}}}
  \newcommand\tauaf{\overline{\tau}^{\mathrm{A}}}
  \newcommand\taudf{\overline{\tau}^{\mathrm{D}}}

  \newcommand\ho{\widehat{\lambda}(\mathcal{G})}
  \newcommand\ha{\widehat{\lambda}_{\mathcal{G}}(m^{\mathrm{A}},0)}
  \newcommand\hd{\widehat{\lambda}_{\mathcal{G}}(m^{\mathrm{A}},m^{\mathrm{D}})}
  \newcommand\hak{\widehat{\lambda}_{\mathcal{G}}(m^{\mathrm{A}}_k,0)}
  \newcommand\hdk{\widehat{\lambda}_{\mathcal{G}}(m^{\mathrm{A}}_k,m^{\mathrm{D}}_k)}
  \newcommand\hato{\widehat{\lambda}_{\mathcal{G}}(\underline{m}^{\mathrm{A}},0)}
  \newcommand\hao{\widehat{\lambda}_{\mathcal{G}}(m^{\mathrm{A}*})}

  \newcommand\haob{\widehat{\lambda}_{\mathcal{G}}(m^{\mathrm{A}2\mathrm{a}*},0)}
  \newcommand\haobb{\widehat{\lambda}_{\mathcal{G}}(m^{\mathrm{A}2\mathrm{b}*},0)}
  \newcommand\haoc{\widehat{\lambda}_{\mathcal{G}}(m^{\mathrm{A}3*},0)}
  \newcommand\hdoc{\widehat{\lambda}_{\mathcal{G}}(m^{\mathrm{A}3*},m^{\mathrm{D}3*})}
  \newcommand\hdoco{\widehat{\lambda}_{\mathcal{G}}(m^{\mathrm{A}},m^{\mathrm{D}3*})}
  
  \newcommand\ina{\delta^{\mathrm{A}}}
  \newcommand\ind{\delta^{\mathrm{D}}}
  \newcommand\inao{\delta^{\mathrm{A}*}}
  \newcommand\indo{\delta^{\mathrm{D}*}}

  \newcommand\inam{\delta^{\mathrm{A}}_l}
  \newcommand\indm{\delta^{\mathrm{D}}_l}
  \newcommand\eam{m^{\mathrm{A}}_l}
  \newcommand\edm{m^{\mathrm{D}}_l}
  \newcommand\go{\mathcal{G}}
  \newcommand\ga{\mathcal{G}^{\mathrm{A}}}
  \newcommand\gd{\mathcal{G}^{\mathrm{D}}}
  \newcommand\gaa{\gamma^{\mathrm{A}}}
  \newcommand\gad{\gamma^{\mathrm{D}}}
  \newcommand\eo{\mathcal{E}}
  \newcommand\ea{m^{\mathrm{A}}}
  \newcommand\ed{m^{\mathrm{D}}}
  \newcommand\eao{m^{\mathrm{A}*}}
  \newcommand\eat{\underline{m}^{\mathrm{A}}}
  \newcommand\eato{\underline{m}^{\mathrm{A}*}}
  \newcommand\edo{m^{\mathrm{D}*}}
  
  \newcommand\eaob{m^{\mathrm{A}2\mathrm{a}*}}
  \newcommand\eaobb{m^{\mathrm{A}2\mathrm{b}*}}
  \newcommand\eaoc{m^{\mathrm{A}3*}}
  \newcommand\edoc{m^{\mathrm{D}3*}}
  
  \newcommand\edob{m^{\mathrm{D}2*}}
  
  \newcommand\dela{\Delta^{\mathrm{A}}}
  \newcommand\deld{\Delta^{\mathrm{D}}}
  \newcommand\delo{\underline{t}}
  \newcommand\delk{\overline{t}}
  
  \newcommand\ua{U^{\mathrm{A}}}
  \newcommand\ud{U^{\mathrm{D}}}

  \newcommand\ba{\beta^{\mathrm{A}}}
  \newcommand\bd{\beta^{\mathrm{D}}}
  
  \newcommand\faa{\hat{U}^{\mathrm{A}1}}
  \newcommand\fab{\hat{U}^{\mathrm{A}2\mathrm{a}}}
  \newcommand\fabb{\hat{U}^{\mathrm{A}2\mathrm{b}}}
  \newcommand\fac{\hat{U}^{\mathrm{A}3}}
  \newcommand\fabo{\hat{U}^{\mathrm{A}2\mathrm{a}*}}
  \newcommand\fabbo{\hat{U}^{\mathrm{A}2\mathrm{b}*}}
  \newcommand\faco{\hat{U}^{\mathrm{A}3*}}
  \newcommand\fdc{\hat{U}^{\mathrm{D}3}}
  \newcommand\fdcdua{\hat{U}_{2}^{\mathrm{D}3}}
  \newcommand\facdua{\hat{U}_{0}^{\mathrm{A}3}}
  \newcommand\fda{\hat{U}^{\mathrm{D}1}}
  \newcommand\fdb{\hat{U}^{\mathrm{D}2\mathrm{a}}}
  \newcommand\fdbb{\hat{U}^{\mathrm{D}2\mathrm{b}}}
  \newcommand\fdco{\hat{U}^{\mathrm{D}3*}}
  \newcommand\ra{\rho^{\mathrm{A}}}
  \newcommand\rd{\rho^{\mathrm{D}}}
  \newcommand\ka{\kappa^{\mathrm{A}}}
  \newcommand\kd{\kappa^{\mathrm{D}}}

  \usepackage{xcolor}
\usepackage{multirow}

\begin{document}
\title{\textcolor{black}{Dynamic Resilient Network Games with Applications to Multi-Agent Consensus}}
\author{Yurid Nugraha, Ahmet Cetinkaya, Tomohisa Hayakawa, Hideaki Ishii, and Quanyan Zhu
\thanks{Yurid Nugraha and Tomohisa Hayakawa are with the Department of  
Systems and Control Engineering, Tokyo Institute of Technology, Tokyo  
152-8552, Japan. {\tt\small{yurid@dsl.sc.e.titech.ac.jp,  
hayakawa@sc.e.titech.ac.jp}}}
\thanks{Ahmet Cetinkaya is with the Information Systems Architecture Science Research Division, National Institute of Informatics, Tokyo 101-8430, Japan. {\tt
\small{cetinkaya@nii.ac.jp}}}
\thanks{Hideaki Ishii is with the Department of Computer Science, Tokyo Insitute of Technology, Yokohama 226-8502,  Japan. {\tt
\small{ishii@c.titech.ac.jp}}}
\thanks{Quanyan Zhu is with the Department of Electrical and Computer Engineering, New York University, Brooklyn, NY 11201, USA. {\tt
\small{quanyan.zhu@nyu.edu}}}
\thanks{This work was supported in the part by the JST CREST Grant No. JPMJCR15K3 and by JST ERATO HASUO Metamathematics for Systems Design Project (No.\ JPMJER1603).}
}
% \author{\IEEEauthorblockN{Michael Shell}
% \IEEEauthorblockA{School of Electrical and\\Computer Engineering\\
% Georgia Institute of Technology\\
% Atlanta, Georgia 30332--0250\\
% Email: http://www.michaelshell.org/contact.html}
% \and
% \IEEEauthorblockN{Homer Simpson}
% \IEEEauthorblockA{Twentieth Century Fox\\
% Email: homer@thesimpsons.com}
% \and
% \IEEEauthorblockN{James Kirk\\ and Montgomery Scott}
% \IEEEauthorblockA{Starfleet Academy\\
% San Francisco, California 96678--2391\\
% Telephone: (800) 555--1212\\
\setcounter{table}{0}
\setcounter{figure}{0}
\maketitle
% \thispagestyle{plain}
% \pagestyle{plain}
\begin{abstract}
A cyber security problem in a networked
system formulated as a resilient graph problem based
on a game-theoretic approach is considered. 
The connectivity of the underlying graph of the network system is
reduced by an attacker who removes some of the edges
whereas the defender attempts to recover them. 
Both players are subject to energy constraints so that their actions are
restricted and cannot be performed continuously. 
For this two-stage game, which is played repeatedly over time, we characterize the optimal strategies for the attacker and the defender in terms of edge connectivity and the number of connected components of
the graph. The resilient graph game is then applied to a multi-agent consensus problem. We study how the attacks and the recovery on the edges affect the consensus process. Finally, we also provide
numerical simulation to illustrate the results.
\end{abstract}
\vspace{-0.3cm}
\section{Introduction}
Multi-agent systems provide a framework for studying
distributed decision-making problems as a number of agents
make local decisions by interacting with each other
over networks \cite{ren,mesba,FB-LNS}.
Due to the rise in the use of general purpose
networks and wireless communication channels for such
systems, cyber security has become a major critical issue \cite{sandb}.
Each agent in the network can be vulnerable to 
various threats initiated by malicious adversaries. One of the common security threats in networked systems
is jamming attacks. The adversary can simply 
transmit interference signals to interrupt
communication among agents.
\textcolor{black}{While jamming attacks against multi-agent systems can
be harmful as it does not require any knowledge of the
systems, the danger level may further increase if 
the attacker is more aware of system parameters.}

Noncooperative game theory approaches are widely used
for addressing security problems including jamming attacks \cite{basar,zhu2}.
Jamming attacks on networked systems were previously
analyzed through game-theoretic approaches. The works \cite{li,li2,gupta} model the activity of jamming and transmitting signals as zero-sum games where the payoff structure of the players is balanced. In \cite{yli,yangj}, the authors consider a Stackelberg game approach, in which the players decide their actions sequentially by following a certain hierarchy.

Multi-agent consensus problems in the presence of such
jamming attacks have been studied in \cite{ali,tesi}. 
The work \cite{kikuchi} introduces a stochastic communication protocol
so that the attackers do not know the exact transmission
times of the agents in advance. Jamming attack models with energy constraints were introduced in \cite{feng,depersis,ahm2,kikuchi} in the context of networked control. These models have been
generalized to further take account of probabilistic
packet losses in \cite{ahm}. \textcolor{black}{In the related studies on resilient consensus, some agents may be attacked by an adversary, making them update their state values in a faulty and even malicious manner; the resilience and robustness in such problems have been discussed in \cite{leblanc,dibaji,guerrero}.} Also, nonmalicious packet losses that can interrupt the communication among agents have been studied in \cite{huang,carli}.

\textcolor{black}{However, in the abovementioned works, optimal strategies for the attackers have not been well addressed. In addition, in those works there is also no defense mechanism to mitigate the attacks and restore the communication so as not to simply wait for the attacks to end.} In this paper, we model the interaction between an
attacker and a defender in a two-player game setting.
The attacker is motivated to disrupt the communication
by attacking individual links while the defender attempts
to recover some or all of them whenever possible.
Both players are constrained in terms of their available
energy for the actions of attacks and recovery.
We extend the problem formulation of \cite{chen}, where the
decision variables 
are limited to the links in the graphs for both players.
\textcolor{black}{In our problem setting, more dynamics are present as the
time intervals for attacking and recovering are to be decided as well.} 

\textcolor{black}{More specifically, in our formulation of resilient graphs, two-stage games are repeatedly played by the attacker and the
defender.} In each attack interval, 
the attacker decides the links and the durations for the attacks.
The attacker's utility depends on the number of connected components of the graph after the attack as well as the remaining energy of the attacker. In response to the attacks, the defender
attempts to recover some of the links that are important for maintaining the connectivity of the graph. Once the attacker ends attacking, the defender also ends recovering since there are no attacks anymore. 
Our study is based on the analysis of the subgame perfect
equilibria of the games, and we use backward induction
to obtain optimal strategies for both players, as in \cite{chen}.

\textcolor{black}{We emphasize that our contribution is the introduction of a game-theoretic framework to jamming attack problems. We follow the attack models dealt with in \cite{feng,depersis,ahm2,kikuchi}, where the energy for communication by the players is under time-varying constraints. Moreover, the defender can overcome the attacker's jamming by sending signals with increased signal-to-interference-plus-noise ratio (SINR); such models are employed in \cite{yli,yangj}. Though the setting is centralized in the sense that both players have control over the networked system, our approach addresses the question on how to design the underlying networks having structures resilient to cyber attacks. As an application of the game problem, we further consider a consensus problem and analyze how the time for reaching consensus is affected by the strategies of the players.}

\begin{figure*}[t]
        \hspace{45pt}
        \psfrag{time}{$t$}
        \psfrag{Edge}{Edge}
        \psfrag{eab}{\scriptsize $e_{12}$}
        \psfrag{eac}{\scriptsize $e_{13}$}
        \psfrag{ebd}{\scriptsize $e_{23}$}
        \psfrag{ebd1}{\scriptsize $e_{34}$}
        \psfrag{ead}{\scriptsize $e_{14}$}
        \psfrag{del10}{\scriptsize $\underline{t}_1$}
        \psfrag{taua1}{\scriptsize $\underline{\tau}^{\mathrm{A}}_1$}
        \psfrag{taud1}{\scriptsize $\underline{\tau}^{\mathrm{D}}_1$}
        \psfrag{taudf1}{\scriptsize $\overline{\tau}^{\mathrm{D}}_1$}
        \psfrag{del20}{\scriptsize $\underline{t}_2$}
        \psfrag{taua2}{\scriptsize $\underline{\tau}^{\mathrm{A}}_2$}
        \psfrag{del21}{\scriptsize $\mathcal{G}^{\mathrm{A}}_1$}
        \psfrag{del22}{\scriptsize $\mathcal{G}^{\mathrm{D}}_1$}
        \psfrag{del01}{\scriptsize $\overline{\tau}^{\mathrm{A}}_1=\overline{t}_1=\underline{t}_2$}
        \psfrag{del02}{\scriptsize $\overline{\tau}^{\mathrm{A}}_2=\overline{t}_2$}
        \psfrag{del22}{\scriptsize $\mathcal{G}^{\mathrm{D}}_1$}
        \psfrag{del23}{\scriptsize $\go$}
        \psfrag{del24}{\scriptsize $\mathcal{G}^{\mathrm{A}}_2$}
        \psfrag{del25}{\scriptsize $\go$}
        \psfrag{eabaa}{\scriptsize $\gaa$}
        \psfrag{eabad}{\scriptsize $\gad$}
        \psfrag{ez}{\scriptsize Attack interval, $\delta^{\mathrm{A}}_1$}
        \psfrag{ez2}{\scriptsize Recovery interval, $\delta^{\mathrm{D}}_1$}
        \psfrag{ez3}{\scriptsize $\delta^{\mathrm{A}}_2$}
        \psfrag{taud2=taudf2}{\scriptsize $\overline{t}_2=\underline{\tau}^{\mathrm{D}}_2=\overline{\tau}^{\mathrm{D}}_2$}
        \includegraphics[width=15 cm]{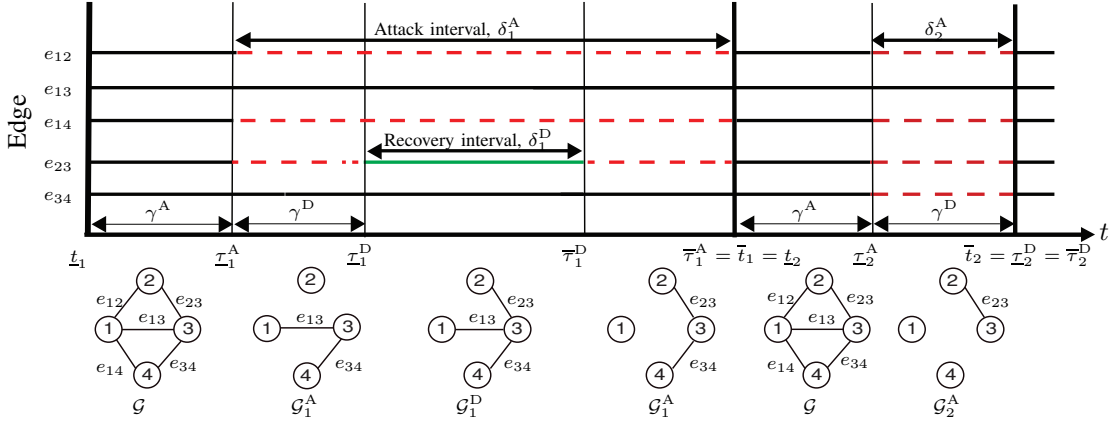}
        \vskip -5pt
        %\caption{Illustration of graph transition. At time interval $[\underline{t}_1,\overline{t}_1]$, the defender recovers one edge $e_{23}$ at $\underline{\tau}^{\mathrm{D}}_1$ and stops recovering at $\overline{\tau}^{\mathrm{D}}_1$. \textcolor{black}{At time interval $[\underline{t}_2,\overline{t}_2]$, the defender cannot recover since the attacker ends jamming at $\underline{\tau}^{\mathrm{D}}_2$, and hence the graph is attacked from $\underline{\tau}^{\mathrm{A}}_2$ to $\overline{t}_2$. Note that the solid lines indicate that the edges are connected, and dashed lines indicate that the edges are disconnected.}}
        \caption{Illustration of graph transition. At time interval $[\underline{t}_1,\overline{t}_1]$, the defender recovers one edge $e_{23}$ at $\underline{\tau}^{\mathrm{D}}_1$ and stops recovering at $\overline{\tau}^{\mathrm{D}}_1$. At time interval $[\underline{t}_2,\overline{t}_2]$, the defender cannot recover since the attacker ends jamming at $\underline{\tau}^{\mathrm{D}}_2$, and hence the graph is represented by $\mathcal{G}^{\mathrm{A}}_2$ from \textcolor{black}{$\underline{\tau}^{\mathrm{A}}_2$} to $\overline{t}_2$. Note that the solid lines indicate that the edges are connected, and dashed lines indicate that the edges are disconnected.}
        \label{fig:a}
        \vspace{-15pt}
    \end{figure*} 

The paper is organized as follows. \textcolor{black}{In Section~II, we
introduce the framework for the resilient
graph game.}
In Section~III, we analyze the subgame perfect equilibria
and characterize the optimal strategies for the players. 
In Section~IV, we apply the obtained results to
a consensus problem for multi-agent systems.
\textcolor{black}{We then provide numerical examples in Section~V and conclude the paper in Section~VI. Finally, all the proofs for our main results are given in the~Appendix. A preliminary version of this paper appeared as \cite{yur}; the scenarios considered there are more restricted as the attacker stops attacking only when running out of energy.}

\vspace{-0.1cm}
\section{Problem Formulation}
    We consider a multi-agent system of $n$ agents with a communication topology described by the undirected graph $\go=(\mathcal{V},\mathcal{E})$. It consists of the set $\mathcal{V}$ of vertices and the set $\mathcal{E} \subseteq \mathcal{V} \times \mathcal{V}$ of edges. The agents are described by the vertices, while the communication links between the agents are represented by the edges. We assume that the underlying, attack-free communication topology $\go$ is connected, i.e., there exists a path connecting every pair of vertices in $\mathcal{V}$.
    
    In this paper, we consider a game between two players, the attacker and the defender, in terms of the communication among the agents. The attacker is an entity capable to block the communication by jamming some targeted links, whereas the defender tries to recover some or all of the attacked links. However, the actions of both players are constrained by the limited energy resources they have.
    
    \textcolor{black}{Our problem setting is centralized in that the attacker and the defender know the conditions of the communication networks at each time and have control over the links individually. That is, the attacker can strategically decide the links to attack while the defender may ask the chosen agents to increase their transmission level to recover their links. As we mentioned in the Introduction, even in such a centralized setting, game-theoretic studies on resilient graphs are very limited. Our game formulation provides insights into networks having resilient structures against adversaries even under a powerful defender having the full knowledge of the system.}
    
    The $k$th game with $k \in \mathbb{N}$ is played in the time interval $[\delo_k,\delk_k]$, which is determined by the players' actions with $\delk_k>\delo_k=\overline{t}_{k-1}$. Initially, at the start time $\delo_k$, there is no attack or recovery, and the underlying graph is $\go$. Then, the attacker may start an attack on certain links, at which point the defender will decide whether to recover some of the attacked links or not. The durations and the links for the attack and the recovery are the action variables. The end time $\delk_k$ is when the attacker and hence the defender stop their actions. The $k$th game may also end after a fixed time duration when no attack occurs. The $(k+1)$th game starts immediately after the $k$th game, that is, $\underline{t}_{k+1}=\delk_k$.
    
    In each time interval $[\delo_k, \delk_k]$, the attacker can start and end attacking, and the defender can start and end recovering at most once. The end of the $k$th time interval $\delk_k$ is specified more concretely later in this section. At the start time $\delo_k$, the active communication links are prescribed by the original edge set $\eo$ for all $k \in \mathbb{N}$. \textcolor{black}{We assume that the attacker fully knows the edge set $\mathcal{E}$}. More specifically, the attacker attacks $\go$ by deleting some of the existing edges $\mathcal{E}^{\mathrm{A}}_k \subseteq \eo$ from time $\taua_k$ until $\tauaf_k$, where $\delo_k < \taua_k \leq \tauaf_k \leq \delk_k$. Consequently, $\go$ is changed to $\ga_k:=(\mathcal{V},\eo \setminus \mathcal{E}^{\mathrm{A}}_k)$ at $\taua_k$. For transmitting jamming signals, the attacker spends some energy in proportion to the attack duration. For the attacker, it is also an option not to make an attack action considering its utility defined later. We define the attack interval as $[\taua_k,\tauaf_k]$ for every $k \in \mathbb{N}$, where the values of $\tauaf_k$ are related to the attacker's energy, as discussed later. If there is no attack in the $k$th time interval, it is understood that $\taua_k=\tauaf_k$.
    
    On the other hand, the defender aims to maintain the connectivity of the graph by recovering some of the edges blocked by the attacker. The defender recovers the edges $\mathcal{E}^{\mathrm{D}}_k$ from time $\taud_k$ until $\taudf_k$, with $\mathcal{E}^{\mathrm{D}}_k \subseteq \mathcal{E}^{\mathrm{A}}_k$ and $\delo_k < \taua_k < \taud_k  \leq \taudf_k \leq \tauaf_k \leq \delk_k$. As soon as the defender starts the recovery action at $\taud_k$, the graph $\ga_k$ is changed to $\gd_k:=(\mathcal{V},(\eo \setminus \mathcal{E}^{\mathrm{A}}_k) \cup \mathcal{E}^{\mathrm{D}}_k))$. By recovering the edges, the defender spends some amount of energy similarly to the attacker. If there is no recovery action due to the absence of the attack action or the decision by the defender, we set $\taud_k=\taudf_k$. We define the recovery interval as $[\taud_k,\taudf_k]$ for every $k \in \mathbb{N}$, where values of $\taudf_k$ are related to the energy of the defender, as discussed later. \textcolor{black}{Once the attacker stops attacking, the attacked edges come back to normal and the graph becomes $\go$ again, which ends the $k$th game and triggers the new $(k+1)$th game.}
    
    In this formulation, we assume that there is a constant dwell time $\gaa > 0$ between the beginning of the $k$th game $\delo_k$ and the beginning of the attack time $\taua_k$. \textcolor{black}{For the defender, there is also a constant dwell time $\gad > 0$ between the beginning of attack time $\taua_k$ and the beginning of recovery time $\taud_k$ unless the attacker ends attacking earlier, i.e., $\tauaf_k<\taud_k$.} Thus, let
    \begin{align} \label{aa}
        \taua_k := \delo_k+\gaa,\quad \taud_k := \min\{\tauaf_k,\taua_k+\gad\}. 
        \end{align}
    The lengths of the attack and the recovery intervals are denoted by $\ina_k$ and $\ind_k$, respectively, with
    \begin{align} \label{ina}
        &\ina_k := \tauaf_k-\taua_k, \quad \ind_k := \taudf_k-\taud_k.
    \end{align}
    \textcolor{black}{The timeline of the attack and the recovery sequences is illustrated in Fig. \ref{fig:a}. It is important to note that two-stage games are repeatedly played by the two players. The equilibrium is thus characterized for each of the two-stage games.}
    
    {\color{black} In the $k$th game, both players attempt to choose the best strategies to maximize their own utility functions defined as how much the agents are connected or disconnected over the time interval [$\delo_k$,$\delk_k$] without foreseeing the future activities. To characterize how much the agents are connected or disconnected in a unified way, we introduce the generalized edge connectivity $\widehat{\lambda}({\mathcal{G}}')$ as an extension of the notion of edge connectivity for the graph $\mathcal{G}'$. It is defined as
    \begin{equation}
        \widehat{\lambda}({\mathcal{G}'}) :=
        \begin{cases} 
        \lambda({\mathcal{G}}'), & \mathrm{if} \ \mathcal{G}' \ \text{is connected}, \\
        -\widetilde{\lambda}({\mathcal{G}}'), & \text{otherwise}, \\
         \end{cases}
    \end{equation}
     where $\lambda({\mathcal{G}}')$ denotes the edge connectivity of the graph $\mathcal{G}'$, i.e., the minimum number of edges required to be removed to make the connected graph $\mathcal{G}'$ disconnected. On the other hand, $\widetilde{\lambda}(\mathcal{G}')$ denotes the minimum number of edges required to make the disconnected graph $\mathcal{G}'$ connected; in this case, there are $\widetilde{\lambda}(\mathcal{G}')+1$ connected components in the disconnected graph $\mathcal{G}'$, since one edge is needed to connect two connected components. {\color{black} Note that a larger positive value of $\widehat{\lambda}$ implies that the graph $\mathcal{G}$ has more links to be removed by the attacker, and a smaller negative value of $\widehat{\lambda}$ indicates that the graph $\mathcal{G}$ requires more links to be recovered by the defender. Since $\ga_k \subseteq \gd_k \subseteq \go$, note that $\widehat{\lambda}(\ga_k) \leq \widehat{\lambda}(\gd_k) \leq \ho $.}
     
     The attacker chooses the optimal edges to attack based on the generalized edge connectivity $\widehat{\lambda}(\mathcal{G})$ of the graph $\go$, and the defender chooses the optimal edges to recover based on the generalized edge connectivity of the graph $\ga_k$. The attacker should strategically choose the edges to jam to reduce $\widehat{\lambda}(\ga_k)$ (making $\ga_k$ more disconnected), and the defender also should choose the edges to efficiently increase $\widehat{\lambda}(\gd_k)$ (making $\gd_k$ more connected). 
     
     Note that for the same number of edges to attack/recover, there may be multiple optimal choices of edges to attack/recover that yield the same values of $\widehat{\lambda}(\ga_k)$ or $\widehat{\lambda}(\gd_k)$. Since we focus on the connectivity of the agents to characterize the utility functions below without specifying particular edges to attack/recover, we define $\widehat{\lambda}_{\mathcal{G}}(\ea_k,\ed_k)$ to represent the generalized edge connectivity of the underlying graph $\mathcal{G}=(\mathcal{V},\mathcal{E})$ with $\ea_k=|\mathcal{E}^{\mathrm{A}}_k|$ edges attacked and $\ed_k=|\mathcal{E}^{\mathrm{D}}_k|$ edges recovered, given by
    \begin{equation}\label{lam}
        \widehat{\lambda}_{\mathcal{G}}(\ea_k,\ed_k) := \min_{\mathcal{E}^{\mathrm{A}}_k:|\mathcal{E}^{\mathrm{A}}_k|=\ea_k} \max_{\mathcal{E}^{\mathrm{D}}_k:|\mathcal{E}^{\mathrm{D}}_k|=\ed_k} \widehat{\lambda}((\mathcal{V},(\mathcal{E}\setminus \mathcal{E}^{\mathrm{A}}_k) \cup \mathcal{E}^{\mathrm{D}}_k)).
    \end{equation}
    For the simple case of $\ed_k=0$, calculating the right-hand side in (\ref{lam}) reduces to the min-cut problem for undirected and unweighted graph $\mathcal{G}$, for which efficient randomized algorithms are available \cite{opeB}. More in general, we can apply the so-called $k$-cut algorithms \cite{ope} by increasing the number $k$ of the connected components. Thus, in principle, the players can obtain the full solution \textit{offline} prior to playing the sequence of the games.
    
    This $\widehat{\lambda}_{\mathcal{G}}$ can be presented as a lower triangular matrix $\widehat{\lambda}_{\mathcal{G}} \in \mathbb{R}^{(|\mathcal{E}|+1) \times (|\mathcal{E}|+1)}$, where $\widehat{\lambda}_{\mathcal{G}}(\ea_k,\ed_k)$ represents the $(\ea_k+1,\ed_k+1)$ entry of the matrix. For example, the matrix $\widehat{\lambda}_{\mathcal{G}}$ for the graph $\mathcal{G}$ in Fig. \ref{fig:a} is given by
    
    \vspace{-5mm}
    \begin{align*} \small \widehat{\lambda}_{\mathcal{G}} =
\begin{bmatrix}
\begin{array}{rrrrrr}
2 & 0 & 0 & 0 & 0 & 0 \\
1 & 2 & 0 & 0 & 0 & 0\\
-1 & 1 & 2 & 0 & 0 & 0 \\
-1 & 1 & 1 & 2 & 0 & 0\\
-2 & -1 & 1 & 1 & 2 & 0\\
-3 & -2 & -1 & 1 & 1 & 2
\end{array}
\end{bmatrix}.
\end{align*}
    In general, the matrix $\widehat{\lambda}_{\mathcal{G}}$ is not Toeplitz, i.e., the values of the $(i,j)$ entries with the same $i-j$ may be different. We also note that the values for the same row/column do not change linearly and that attacking/recovering more number of edges does not necessarily change the graph connectivity.
     
    The strategies of the attacker and the defender are in terms of $(\ea_k, \ina_k)$ and $(\ed_k, \ind_k)$, respectively. For the game of the $k$th time interval $[\delo_k,\delk_k]$, we define the utility function $\ua$ of the attacker as 
    \begin{align} \label{ua}
    \ua&((\ea_k,\ina_k), (\ed_k,\ind_k)) \notag \\ := & -\hak(\ina_k-\ind_k)  - \hdk \ind_k - \ba\ea_k\ina_k,
    \end{align}
    where $\ba>0$ is the attacker's cost to remove one edge per time unit. Similarly, define the utility function $\ud$ of the defender as
    \begin{align} \label{ud}
    \ud&((\ea_k,\ina_k),(\ed_k,\ind_k)) \notag \\ := & \ \hak(\ina_k-\ind_k) + \hdk \ind_k - \bd\ed_k\ind_k,
    \end{align}
    where $\bd>0$ is the defender's cost to recover one edge per time unit. Note that the utility function (\ref{ua}) represents the total generalized edge connectivity (with the negative sign) for the attacker over the game horizon $[\taua_k,\delk_k]$ plus the cost for jamming $\ea_k$ number of communication links. Similarly, (\ref{ud}) represents the total generalized edge connectivity for the defender over the game horizon $[\taua_k,\delk_k]$ plus the cost for recovering $\ed_k$ number of communication links.}
    
    If the attacker decides to attack at least one edge, then the game ends at $\tauaf_k$. Otherwise, the game ends at $\delo_k+\gaa+\gad$. In other words, the end time $\delk_k$ of the $k$th game is 
    \begin{equation} \label{delt}
        \delk_k :=
        \begin{cases} 
        \tauaf_k, & \mathrm{if} \ \ea_k > 0, \\
       \delo_k+\gaa+\gad, & \mathrm{otherwise}. \\
         \end{cases}
    \end{equation}
    \textcolor{black}{According to the utility functions (\ref{ua}) and (\ref{ud}), there is a case where the defender stops recovering $\ed_k$ number of links before the game ends while the attacker keeps sending jamming signals to $\ea_k$ number of links. In this case, the graph changes back to $\ga_k$ at $\taudf_k$, with generalized edge connectivity $\hak$. Therefore, in $[\taudf_k,\delk_k]$, the utilities of both players in ($\ref{ua}$) and ($\ref{ud}$) are computed based on $\hak$.}
    
    The players cannot keep sending signals for very long durations due to energy constraints. We follow the approach in \cite{kikuchi} to model such energy constraints. {\color{black} The total energy used by player $p\in\{\mathrm{A},\mathrm{D}\}$ must satisfy
    \begin{equation} \label{a}
        \sum_{l=1}^{k-1} \beta^p m^p_l \delta^p_l +\beta^p m^p_k(t-\underline{\tau}^p_k) \leq \kappa^p + \rho^p t,
    \end{equation}
     for any time $t \in [\underline{\tau}^p_{k},\underline{\tau}^p_{k+1}]$, with $\kappa^p>0$, $ \rho^p \in (0,1)$, $\beta^p > \rho^p$, and $k \in \mathbb{N}$. Note that $\kappa^p$ denotes the initial energy that player $p$ has, and $\rho^p$ denotes the recharge rate of energy for player $p$. The left-hand side of (\ref{a}) represents energy consumed by player $p$ up to time $t$ and is affected by the number of attacked/recovered edges and the attack/recovery durations from the first game until the $k$th game. The right-hand side represents the total available energy, dictated by the parameters $\kappa^p$ and $\rho^p$. In this paper, we assume that each player knows all parameters of the other player, including $\kappa^p$ and $\rho^p$. 
    
   Under this problem formulation, if player $p$ keeps sending jamming/recovering signals starting at time $\underline{\tau}^p_k$ until running out of energy, then from (\ref{a}) we obtain an explicit expression for the maximum interval $\Delta^p_k$ on the time duration $\delta^p_k$ when player $p$ completes the attack/recovery as }
   {\color{black} \begin{equation} \label{en.a}
        \Delta^p_k(m^p_k) :=  \frac{\kappa^p+\rho^p\underline{\tau}^p_k-\sum_{l=1}^{k-1} \beta^p m^p_l \delta^p_l}{\beta^p m^p_k-\rho^p}.
    \end{equation}
    }

    \textcolor{black}{We formulate the $k$th game as a two-stage game where the attacker first attacks and then the defender makes recoveries. It should be noted that each game is played independently at time $\delo_k$ and the strategies of the players will depend on their energy level at that point. It is, however, noted that there would be a preceding stage, which is implicit in our formulation; this stage is related to the design of the network structure of the underlying graph $\mathcal{G}$. The underlying graph is assumed to be given in this paper, but clearly affects the game as it is the default network at the start of each game. In this respect, our formulation will be useful in finding resilient networks under hostile environments.}

    We seek the subgame perfect equilibrium of the $k$th game as in \cite{chen}. To this end, one needs to divide the game into some subgames. The equilibrium must be optimal in every subgame. The defender's game is formulated as a subgame of the attacker's game. Therefore, the attacker also maximizes the defender's utility function to obtain the defender's best strategy given the attacker's strategy, and uses the defender's best strategy to formulate the best strategy for the attacker. \textcolor{black}{To obtain the optimal strategy for each player, \textit{backward induction} is used in each $k$th game consisting of two-stage decision-making levels corresponding to the attack and recovery sequences. This two-stage game is played independently at time $\underline{t}_k$. Notice that since the maximum durations of attacks and recoveries in (\ref{en.a}) are affected by the players' strategies in the past games, the players' strategies in the $k$th game are influenced by their strategies in the previous games and therefore the players' optimal strategies can be different in each game.} 
    
    In the time interval $[\delo_k,\delk_k]$, given the attacker's strategy $(\ea_k,\ina_k)$, the defender decides the strategy as
    \begin{align}
    (\edo_k & (\ea_k,\ina_k),\indo_k(\ea_k,\ina_k)) \notag \\ & \in \argmax_{(\ed_k,\ind_k)}  \ud((\ea_k,\ina_k),(\ed_k,\ind_k)), \label{st1}
    \end{align}
    with $\ed_k$ and $\ind_k$ depending on $\ea_k$ and $\ina_k$. \textcolor{black}{Likewise, given the initial graph $\go$, the attacker decides the strategy as}
    \begin{align}
    (&\eao_k,\inao_k) \notag  \notag \\ & \in \argmax_{(\ea_k,\ina_k)} \ua  ((\ea_k,\ina_k),(\edo_k(\ea_k,\ina_k), 
    \indo_k(\ea_k,\ina_k))). \label{st2}
    \end{align}
    
    We study the subgame perfect equilibrium and seek pairs $(\ea_k,\ina_k)$ and $(\ed_k,\ind_k)$ such that $(\ed_k,\ind_k)$ is the best response to $(\ea_k,\ina_k)$. The combination of strategies $((\ea_k,\ina_k),(\ed_k,\ind_k))$ that follow the subgame perfect equilibrium principle is called the \textit{optimal combined strategy}. A tie-break condition happens if the players have multiple options for the choices on which edges to attack or recover, and those edges yield the same values of the utility functions. In this case, we suppose that the players choose more edges to attack or recover.
\vspace{-0.1cm}
\section{Game Analysis}
    \textcolor{black}{In this section, we discuss the subgame perfect equilibrium formulation and the characteristics of the players on one game interval. Hence, in this section we remove the subscript $k$ from all variables. We assume that the maximum attack/recovery durations $\dela(\ea)>0$ and $\deld(\ed)>0$ are given. For simplicity of notation, we omit the variable $\ea$ (resp., $\ed$) for the presentation of $\dela$ (resp., $\deld$) in this section.}
    \vspace{-0.3cm}
\subsection{Brief Summary of the Results}
     We first provide a summary of the results. To characterize the optimal strategies, from the sequence of actions by the attacker and the defender described in the previous section, we categorize the possible combinations of generalized edge connectivities $\ho$, $\ha$, and $\hd$ into three cases shown in Table \ref{tab:my_label}. Note that these cases cover all the possible combinations of the actions by both players. \textcolor{black}{Since $\ed \leq \ea$, it is impossible to have $\ha=\ho$ and $\hd>\ha$. Also, note that since by definition $\ed \geq 0$, condition $\hd<\ha$ cannot be fulfilled.} Furthermore, even if the attacker attacks some edges of $\eo$, there is a possibility that the edge connectivity does not change, as in Case 1. The same remark applies to the recovery action. As a result, there are four possible optimal combined strategies that are derived from the three cases in Table~\ref{tab:my_label}. A summary of the results of the optimal strategies is shown in Table \ref{tab:tab2}. \textcolor{black}{Note that it may be optimal for the attacker to continue attacking even after the recovery finishes, since the attacker gets higher utility in $[\taudf,\tauaf]$.}

    \begin{table}
        \caption{Possible cases of attack and recovery actions}
        \renewcommand{\arraystretch}{1.4}
        \label{tab:my_label}
        \centering
        \vskip -5pt
        \begin{tabular}{|c|c|c|}
        \hline
            Case & $\ha$ & $\hd$  \\
            \hline
            1 & $\ha=\ho$ & $\hd=\ha$ \\
            \hline
            2 & $\ha < \ho$ & $\hd = \ha$ \\
            \hline
            3 & $\ha < \ho$ & $\hd > \ha$ \\
            \hline
            \end{tabular}
        \vskip -5pt
    \end{table}

 \begin{table}
        \caption{Optimal combined strategy candidates}
        \vskip -5pt 
        \renewcommand{\arraystretch}{1.2}
        \label{tab:tab2}
        \centering
        \begin{tabular}{|c|l|}
        \hline
            Comb. & \multirow{2}{*}{Action}  \\
            Str. &   \\
            \hline
            1 & Attacker: No attack \\ & Defender: No need to recover  \\
            \hline
            2a & Attacker: Attacks the optimal edges for $\dela$ duration \\ & Defender: No recovery \\
            \hline
            2b & Attacker: Attacks the optimal edges until $\taud$ \\ & Defender: No chance to recover\\
            \hline
            3 & Attacker: Attacks the optimal edges for $\dela$ duration \\  & Defender: Recovers the optimal edges for \\ &  \hspace*{32pt} $\min\{\deld,\dela+\taua-\taud\}$ duration\\
            \hline
            \end{tabular}
        \vskip -10pt    
    \end{table}   
   \vspace{-0.3cm}
   \subsection{Subgame Perfect Equilibrium Analysis}
   In this subsection, we analyze the subgame perfect equilibrium of the system. From the sequence of actions, we obtain several cases that might happen and seek the equilibrium in each case, i.e., the candidate optimal strategies of the system. Then, we seek the optimal strategy among the candidate strategies by using backward induction.
   
    \subsubsection{Subgame Perfect Equilibrium Analysis in Each Case}
    \textcolor{black}{From the problem formulation, since $\ho \geq \hd \geq \ha$, we obtain three cases based on the combinations of $\ho$, $\ha$, and $\hd$, as shown in Table \ref{tab:my_label}. We analyze the subgame perfect equilibrium for the time interval $[\delo,\delk]$ in each case. The results in terms of links and durations of the optimal combined strategy candidates are summarized in Table~\ref{tab:reg1}.} 
    
    \indent \textbf{Case 1:} \textcolor{black}{In this case, we show that the optimal strategy for the players are not to recover any edge, i.e., $\eao,\edo=0$.} By Table~\ref{tab:my_label}, the utility function in (\ref{ud}) of the defender becomes
    \begin{equation} \label{sit1}
        {\color{black}\ud((\ea,\ina),(\ed,\ind)) = \ho\ina-\bd\ed\ind.}
        \end{equation} 
        
    {\color{black} Furthermore, because the defender receives no reward by recovering any link, the optimal strategy for the defender is $\edo=0$ and $\indo=0$, resulting in
        \begin{equation} \label{p1df}
             \ud((\ea,\ina),(\edo,\indo)) = \ho\ina.
        \end{equation}
    This strategy $\edo=0$ and $\indo=0$ for the defender is named \textbf{Strategy D1} (see Table \ref{tab:reg1}).
    
        \begin{table}[t]
            \renewcommand{\arraystretch}{1.5}
            \vspace{10pt}
            \caption{\textcolor{black}{Links and durations of the optimal combined strategy candidates}}
            \vskip -10pt
            \label{tab:reg1}
            \begin{center}
            \begin{tabular}{|c||c|c|c|c|c|c|}
            \hline
            Comb.  & Att. & \multirow{2}{*}{$\eao$} & \multirow{2}{*}{$\inao$} & Def.& \multirow{2}{*}{$\edo$} & \multirow{2}{*}{$\indo$} \vspace{-5pt}\\
            Str. & Str. & & & Str. & & \\
            \hhline{|=||=|=|=|=|=|=|}
            1 & A1 & $0$ & $0$ & \multirow{3}{*}{D1} & \multirow{3}{*}{$0$} & \multirow{3}{*}{$0$} \\
            \cline{1-4}
            2a & A2a & $\eaob$ & $\dela(\eaob)$& & &   \\
            \cline{1-4}
            2b & A2b & $\eaobb$ & $\taud-\taua$ & & & \\
            \hline
            \multirow{2}{*}{3} & \multirow{2}{*}{A3} & \multirow{2}{*}{$\eaoc$} & \multirow{2}{*}{$\dela(\eaoc)$} & \multirow{2}{*}{D3} & $\edoc$ & \multirow{2}{*}{$\xi$} \\
             &  & & & & $(\eaoc)$ & \\
            \hline
            \end{tabular}
            \vskip -15pt
        \end{center}
        \vskip -15pt
    \end{table}

    Likewise, for the attacker, the utility function in (\ref{ua}) becomes
        \begin{equation} \label{p1a}
            \ua((\ea,\ina),(\edo,\indo)) = (-\ho-\ba\ea)\ina.
        \end{equation}}
    {\color{black} It is then clear that the optimal strategy for the attacker is $\eao=0$ and $\inao=0$. As a result, the utility functions in Case 1 are given by
        \begin{align} 
             \ua((\eao,\inao),(\edo,\indo)) = 0  &=: \faa \label{p1af},\\
             \ud((\eao,\inao),(\edo,\indo)) = 0 &=: \fda .
        \end{align}
        From (\ref{delt}), because $\ea=\ed=0$, it follows that the game ends at $\delk=\delo+\gaa+\gad$. This optimal strategy candidate $\ea,\ina=0$ for the attacker is classified as \textbf{Strategy A1}. In this case, the optimal combined strategy corresponding to $((\eao,\inao),(\edo,\indo))$ is then labelled as \textbf{Combined Strategy~1 := (Strategy A1, Strategy D1)}.}
        
    \textbf{Case 2:} {\color{black} In this case, we show that the attacker's optimal strategy is to attack until running out of energy, whereas the optimal strategy for the defender is not to recover any edge. 
    
    Similarly with the analysis in Case 1, because $\hd=\ha$, the utility function of the defender with $\edo, \indo=0$ as in (\ref{p1df}) is given by
        \begin{align} \label{p3f}
             \ud((\ea,\ina),(\edo,\indo)) &= \ha\ina.
        \end{align}
    For the attacker, from (\ref{ua}) with $\ind=0$, we have
        \begin{equation}
              \ua((\ea,\ina),(\edo,\indo)) = (-\ha-\ba\ea)\ina.
        \end{equation}
    If $-\ha-\ba\ea>0$, the attacker maximizes $\ina$, by attacking as long as possible. Hence, $\ina=\dela$, and 
        \begin{align}
            &\ua((\ea,\inao),(\edo,\indo)) \notag  \\
             & = (-\ha-\ba\ea)\dela =: \fab (\ea). \label{hoa}
        \end{align}
    Now we only need to choose $\ea$, as $\ina$ is already determined. Specifically, we search for $\eaob$, which denotes the optimal $\ea$. This is done by maximizing the simplified utility function $\fab(\ea)$ in (\ref{hoa}), resulting in
    \begin{align}
         \eaob & \in \argmax_{\ea > 0} \fab(\ea).\label{ex}
    \end{align}
    Note that with this strategy, (\ref{p3f}) becomes
    \begin{align}
        \ud((\eao,\inao),(\edo,\indo))&=\haob\dela =: \fdb.
    \end{align}}
    The attacker's strategy in this case is specified as \textbf{Strategy~A2a}, which is $\ea=\eaob$ and $\ina=\dela$. This combination of strategies of $((\eao,\inao),(\edo,\indo))$ is labelled as \textbf{Combined Strategy~2a := (Strategy A2a, Strategy D1)}.

    \textbf{Case 3:} In this case, we show that the optimal strategy for the attacker is to attack the optimal edges until running out of energy or to attack until the defender starts to recover, whereas the optimal strategy for the defender is to recover the optimal edges until the defender runs out of energy or the attacker ends attacking. {\color{black} In this case, by Table \ref{tab:my_label}, the generalized edge connectivities satisfy $\ho \geq \hd > \ha$. 
    
    From (\ref{ud}), the defender's utility function can be written as
        \begin{align}
             \ud((\ea,\ina),(\ed,\ind)) & = \phi \ind +\ha \ina, \label{tai2}
        \end{align}
    with $\phi:=(\hd-\ha-\bd\ed)$ for simplicity.
    Since $\ha<\hd$, in order to maximize the term $\phi\ind$, the defender recovers $\ed$ links as long as possible if $\phi\geq 0$, so that $\taudf=\min \{\deld+\taud,\tauaf\}$. Alternatively, if $\phi < 0$, then the defender should not recover. It follows that the utility function of the defender becomes
        \begin{align} \label{tai}
             \ud&((\ea,\ina),(\ed,\min \{\deld,\tauaf-\taud\})) \notag \\ & = \phi (\min \{\deld,\tauaf-\taud\}) +\ha\ina.
        \end{align}
        
    Since the attacker is able to attack for $\dela$, we divide the analysis for this case into two parts: (i) the attacker ends attacking before $\deld+\taud$, and (ii) the attacker ends attacking after $\deld+\taud$. 
        
    (i) In this case, the attacker ends the game before the defender finishes the recovery attempt that would have lasted for $\deld$ units of time. However, since the attacker ends the game earlier, the recovery duration is only $\tauaf-\taud$ units of time. Thus, we have $\taudf=\tauaf=\delk$, and the attacker's utility function in (\ref{ua}) can be stated as
        \begin{align} \label{p4ax}
             \ua(&(\ea,\ina),(\ed,(\tauaf-\taud))) \notag \\ \quad & = (-\ha-\ba\ea) (\taud-\taua) \notag \\ & \quad + (-\hd-\ba\ea) (\tauaf-\taud).
        \end{align}}
    (ii) In this case, the attacker ends the game after the defender finishes the recovery attempt. {\color{black} Hence, $\taudf=\deld+\taud$, where the utility function for the attacker keeps the form as in ($\ref{ua}$). 
    
    \textbf{Combined Strategy 3:} From (i) and (ii) above, one of the obvious choices for the attacker is to attack for $\dela$ duration. Depending on the value of $\dela$, the attacker can end attacking before or after $\deld+\taud$. If the attacker ends attacking before $\deld+\taud$, then $\delk=\taudf=\dela+\taua$. Otherwise, the defender recovers for $\deld$, and $\deld+\taud<\delk=\dela+\taua$. Hence, we can rewrite (\ref{tai}) as
       \begin{align} \label{ho3}
             \ud((\ea,\dela),(\ed,\xi)) & = \phi \xi +\ha\ina \notag \\ & =: \fdc(\ea,\ed),
        \end{align}
    with 
    \begin{align} \label{xi}
             \xi:=\min\{\deld,\dela+\taua-\taud\}.
    \end{align}
    Then the optimal number of edges to be recovered for given $\ea$ is obtained by
     \begin{equation} \label{ex4a}
            \edoc(\ea) \in 
            {\argmax_{\ed > 0}} \ \fdc(\ea,\ed).
    \end{equation} 
    The utility function of the attacker can be rewritten as
        \begin{align} \label{hoa3}
           \ua&((\ea,\inao),(\edo,\indo)) \notag\\
            &=-\ha (\dela-\xi) - \hdoco \xi - \ba\ea\dela \notag \\ & =: \fac(\ea).
        \end{align}
    The attacker looks for the optimal number of edges $\eaoc$ by maximizing the simplified utility function $\fac(\ea)$. Specifically,
       \begin{align} 
         \eaoc & \in \argmax_{\ea > 0} \fac(\ea). \label{ex3}
        \end{align}
    Note that to obtain $\eaoc$, the attacker needs to obtain $\edoc$ first. Hence, the attacker solves the maximization problem in (\ref{ex4a}) beforehand to obtain $\edoc(\ea)$.} This strategy for the attacker is named as \textbf{Strategy A3}.
   
    {\color{black} Finally, after the attacker obtains $\eaoc$, the defender searches for $\edoc$, based on $\fdc(\eaoc,\ed)$ in (\ref{ho3}), as 
       \begin{equation} \label{ex4}
        \begin{split}
            \edoc(\eaoc) \in 
            {\argmax_{\ed > 0}} \ \fdc(\eaoc,\ed).
        \end{split}
    \end{equation}
    This strategy $\ed=\edoc(\eaoc)$, $\ind=\xi$ for the defender is labelled as \textbf{Strategy D3.}}
    We call this combined strategy as \textbf{Combined Strategy 3 := (Strategy A3, Strategy D3)}.
    
    {\color{black} 
    
    \textbf{Combined Strategy 2b:} Another choice of the attacker is to end attacking at $\taud$, which is preferred if $-\hd -\ba\ea<0$ (from the second term of (\ref{p4ax})), i.e., the cost of attacking is too high at interval $[\taud,\tauaf]$. Since the attacker ends attacking at $\taud$, the defender cannot recover any edge (Strategy D1), i.e., $\ed=0$ and $\ind=0$. Consequently, the attacker's utility function becomes
        \begin{align} \label{hoa4}
             &\ua((\ea,\inao),(\edo,\indo)) \notag \\ & = (-\ha-\ba\ea) (\taud-\taua) =: \fabb (\ea).
        \end{align}
    As in the previous strategy, the attacker looks for the optimal number of edges $\eaobb$ by maximizing the simplified utility function $\fabb(\ea)$. Specifically,
       \begin{align} 
         \eaobb & \in \argmax_{\ea > 0} \fabb(\ea). \label{ex5}
        \end{align}   
    Strategy $\ea=\eaobb$ and $\ina=\taud-\taua$ for the attacker is specified as \textbf{Strategy A2b}.
    Note that with this strategy, utility function in (\ref{tai2}) becomes
    \begin{align}
        \ud((\eao,\inao),(\edo,\indo))&=\haobb\dela =: \fdbb.
    \end{align}
    As $\ha<\ho$ and $\hd=\ha$, this optimal strategy of $((\eao,\inao),(\edo,\indo))$ is named as \textbf{Combined Strategy~2b := (Strategy~A2b, Strategy~D1)}. }

    \subsubsection{Subgame Perfect Equilibrium Analysis of All Cases}
    Here, we discuss the subgame perfect equilibrium analysis of the system among all cases. Specifically, we find the strategy that yields the maximum utility out of the four possible combined strategies described in Section~III.B.1, in accordance with the subgame perfect equilibrium principle. \textcolor{black}{This is done by applying the backward induction method to the maximum values of the simplified utility functions $\faa$, $\fabo :=\fab(\eaob)$, $\fabbo=\fabb(\eaobb)$, $\faco := \fac(\eaoc)$, $\fda$, $\fdb$, $\fdbb$, and $\fdco :=\fdc(\eaoc,\edoc(\eaoc))$.} 
    
    We first state properties of utility functions in some strategies. In Lemma \ref{30}, we state that the attacker's utility without recovery is always higher than the one with recovery by the defender, for the same $\ea$ and $\ina$. {\color{black} Lemmas \ref{31} and \ref{32} characterize the properties of $\fabo$, $\fabbo$, and $\faco$ in terms of their values relative to others.}
    \vspace{-0.1cm}
    \begin{lemma}\label{30}
    For all possible combinations of $\ed$ and $\ind$, it holds $\ua((\ea,\ina),(0,0)) \geq \ua((\ea,\ina),(\ed,\ind))$.
    \end{lemma}
    \vspace{-0.1cm}
    \vspace{-0.1cm}
    \begin{lemma} \label{31}
    For any possible $\eaob$ and $\eaoc$, it follows that $\fabo \geq \faco$. 
    \end{lemma}
    \vspace{-0.1cm}
\vspace{-0.1cm}
    \begin{lemma} \label{32}
    $\fabo$ has the same sign with $\fabbo$. Also, $\fabo \geq \fabbo$ if $\fabo>0$.
    \end{lemma}
    \vspace{-0.1cm}

    We are now ready to state the main result of this paper. {\color{black} Since $\widehat{\lambda}_{\mathcal{G}}(\ea,\ed)$ is a nonlinear function of $\ea$ and $\ed$ and its particular form depends on the underlying graph $\mathcal{G}$, the utility functions cannot be represented as simple functions of the action and energy variables except for certain cases. For this reason, we present our general result in terms of the functions $\hat{U}^*$. In particular, we use $\facdua$ and $\eato$ defined by 
    \begin{align}
    \facdua &:= \max_{\ea \in \mathcal{M}} \fab(\ea), \\
    \eato &\in \argmax_{\ea \in \mathcal{M}} \fab(\ea) \label{38},
    \end{align}
    where $\mathcal{M}:=\{\underline{m}^{\mathrm{A}} \in \{0,|\mathcal{E}|\}:\widehat{\lambda}_{\mathcal{G}}(\eat,m^{\mathrm{D3}*})-\hato-\bd\edoc<0\}$. Furthermore, we let $\fdcdua :=\fdc(\eaob,\edoc(\eaob)).$
\vspace{-0.1cm}
    \begin{theorem}\label{33}
    The subgame perfect equilibrium of the $k$th game in the time interval $[\delo,\delk]$ satisfies the following:
    \begin{enumerate}
    \item Combined Strategy 1 is optimal if $\fabo<0$.
    \item Combined Strategy 2a is optimal if $\fabo \geq 0$ and
    \begin{enumerate}
        \item $\fdcdua<\fdb$, or
        \item $\fdcdua\geq\fdb$ and
        \begin{enumerate}
            \item $\faco < \fabbo$ and $\facdua>\fabbo$, or
            \item $\faco \geq \fabbo$ and  $\facdua > \faco$.
        \end{enumerate}
    \end{enumerate}
    In these cases \textrm{(a)} and \textrm{(b)} above, the optimal number of edges $\eao$ for the attacker to attack are $\eaob$ and $\eato$, respectively .
    \item Combined Strategy 2b is optimal if $\fabo\geq 0$, $\fdcdua\geq\fdb$, $\fabbo>\faco$, and $\fabbo > \facdua$.
    \item Combined Strategy 3 is optimal if $\faco \geq \fabbo \geq 0$, $\fdcdua \geq \fdb$, and $\facdua \leq \faco$.
    \end{enumerate}
    The combined strategies above cover all possible cases. 
   \end{theorem}}\vspace{-0.1cm}

    \begin{figure}
        \centering
        \psfrag{A1}{\scriptsize \textcolor{black}{ $(\eao,\inao)=(0,0)$}}
        \psfrag{A2A}{\scriptsize $(\eaob,\dela)$}
        \psfrag{A3}{\scriptsize $(\eaoc,\dela)$}
        \psfrag{A4}{\scriptsize $(\edo,\indo)=(0,0)$}
        \psfrag{A5}{\scriptsize \textcolor{black}{$(0,0)$}}
        \psfrag{A6}{\scriptsize \textcolor{black}{$(0,0)$}}
        \psfrag{A7}{\scriptsize $(\edoc(\eaob),\xi)$}
        \psfrag{A8}{\scriptsize \textcolor{black}{$(0,0)$}}
        \psfrag{A9}{\scriptsize $(\edoc(\eaoc),\xi)$}
        \psfrag{A10}{\scriptsize $(\faa,\fda)$}
        \psfrag{A11}{\scriptsize $(\fabo,\fdb)$}
        \psfrag{A12}{\scriptsize $(\fabo,\fdb)$}
        \psfrag{A13}{\scriptsize $(\cdot,\fdcdua)$}
        \psfrag{A14}{\scriptsize $(\cdot,\haoc\dela)$}
        \psfrag{A15}{\scriptsize $(\faco,\fdco)$}
        \psfrag{A16}{\scriptsize $(\eaobb,(\taud-\taua))$}
        \psfrag{A17}{\scriptsize $(0,0)$}
        \psfrag{A18}{\scriptsize $(\fabbo,\fdbb)$}       
        \hspace{-2cm}
        \includegraphics[scale=0.7]{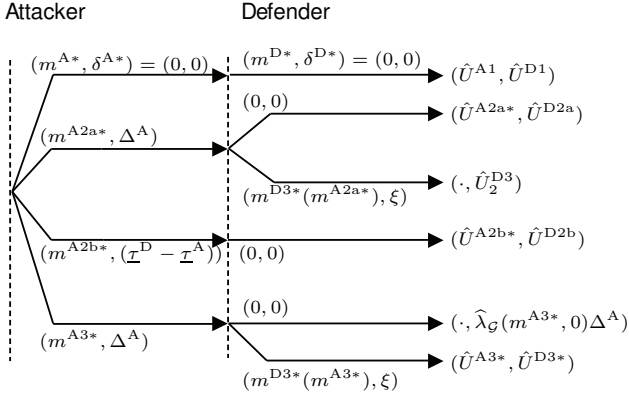}
        \caption{Illustration of possible optimal strategies. Arrows that represent possible actions of the attacker and the defender lead to pairs of utilities obtained under those actions. {\color{black} The dot in the attacker's utilities in $(\cdot,\fdcdua)$ and $(\cdot,\haoc\dela)$ means that those utilities are not considered to find the optimal strategy.}}
        \label{fig:my_label3}
        \vskip -10pt
    \end{figure}

\begin{table*}
	\centering
	\begin{minipage}{1.2\columnwidth}
		\centering
		\hspace{-5cm}
            \renewcommand{\arraystretch}{1.6}
            \captionof{table}{\textcolor{black}{Characterization of the optimal strategy of all cases}}
            \vspace{-5pt}
            \label{tab:heh}
            \begin{tabular}{|c|c|c||c|c|}
            \hline
            \multicolumn{3}{|c||}{Conditions} & $\fdcdua < \fdb$& $\fdcdua \geq \fdb$\\
            \hhline{|=|=|=||=|=|}
            \multirow{4}{*}{$\fabo \geq 0$} & \multirow{2}{*}{$\faco < \fabbo$} & $\facdua \geq \fabbo$ & \multirow{4}{*}{Comb. Str. 2a} &  Comb. Str. 2a\\
            \cline{3-3} \cline{5-5}
             &  & $\facdua < \fabbo$ & & Comb. Str. 2b\\
            \cline{2-3}\cline{5-5}
              & \multirow{2}{*}{$\faco \geq \fabbo$} & $\facdua > \faco$ &  & Comb. Str. 2a\\
            \cline{3-3}\cline{5-5}
             &   & $\facdua \leq \faco$ &  & Comb. Str. 3\\
            \hline
             \multicolumn{3}{|c||}{$\fabo < 0$} & \multicolumn{2}{|c|}{Comb. Str. 1}  \\
            \hline
            \end{tabular}
	\end{minipage}%
	\hfill
	\begin{minipage}{.8\columnwidth}
		\centering
		%\hspace{-2cm}
		\psfrag{A1}{\scriptsize \textcolor{black}{ $\ba$}}
        \psfrag{ba4}{\scriptsize $\bd$}
        \psfrag{ba5}{\scriptsize $\ra$}
        \psfrag{ba6}{\scriptsize $\rd$}
        \psfrag{ba1}{\scriptsize $1$}
        \psfrag{ba3}{\scriptsize $2$}
        \psfrag{ba2}{\scriptsize $1-\frac{2\xi}{\dela-\taud+\taua}$}
		\includegraphics[scale=0.4]{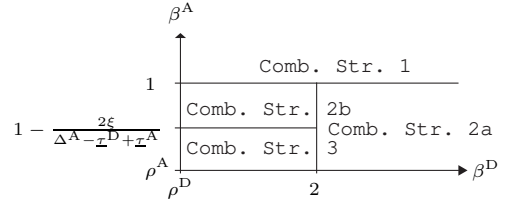}
		\captionof{figure}{\textcolor{black}{Optimal strategies of all cases for $n=2$}}
		 \label{fig:fign2}
	\end{minipage}
	\vskip -10pt
\end{table*}

   Possible optimal strategies for both players are illustrated in Fig. \ref{fig:my_label3}. \textcolor{black}{Moreover, combinations of the conditions of the possible optimal strategies in all cases are shown in Table~\ref{tab:heh}. We also note that even if the unit costs $\ba$ and $\bd$ for attacking/recovering one edge per time depend on edges, the procedure to find the optimal combined strategies as in Theorem~\ref{33} does not change.}
    
    \textcolor{black}{From the optimal strategies in Theorem \ref{33}, we can state some corollaries about the effects of the uniform cost $\ba$ and $\bd$ to the optimal strategy as follows.} It is interesting to note that the critical values of $\ba$ and $\bd$ are different.
    \begin{corollary} \label{corb}
    The optimal strategy for the defender is not to recover if $\bd > 2$.
    \end{corollary} \vskip 2pt
    \vspace{-0.1cm}

    \begin{corollary}\label{cora}
    The optimal strategy for the attacker is not to attack if $\ba > 1$. {\color{black} Also, under the optimal strategy, if the attacker attacks (i.e., $\ea, \ina > 0$), then $\ga$ always becomes disconnected.}
    \end{corollary} \vskip 2pt

    \begin{remark}
    {\color{black} If $\ha < 0$ (i.e., $\ga$ is disconnected), then in order to make $\hd$ larger, the defender can reduce the number of connected components by adding links until the graph becomes connected ($\hd>0$). The minimum number of edges to add in order to achieve certain $\hd$ in a disconnected $\ga$ is given by
    \begin{align} \label{br}
        \ed = \ & \hd-\ha, \notag \\ 
        & \ \mathrm{for} \ \hd < 0,\ha < 0.
    \end{align}}
    \end{remark}
    
    {\color{black}\vspace{-0.5cm} To provide a more explicit relation between optimal strategies and attack/recovery parameters, we present a result for a simple case. It allows us to determine the equilibrium based on the cost and action durations. To this end, we consider a graph with $n=2$ and $|\mathcal{E}| \ = 1$. In this setup, both players can only attack/recover one edge. Based on the results in Theorem~\ref{33}, the optimal combined strategy can be stated as follows.} 
    \begin{proposition}\label{37}
    \textcolor{black}{The optimal combined strategy of the players with $n=2$ is given by}
     \begin{enumerate}
    \item \textcolor{black}{Combined Strategy 1 if $\ba > 1$;}
    \item \textcolor{black}{Combined Strategy 2a if $\ba \leq 1$ and $\bd>2$;}
    \item \textcolor{black}{Combined Strategy 2b if $1-\frac{2\xi}{\dela-\taud+\taua}<\ba\leq 1$ and $\bd\leq2$;}
    \item \textcolor{black}{Combined Strategy 3 if $\ba \leq 1-\frac{2\xi}{\dela-\taud+\taua}$ and $\bd \leq 2$.}
     \end{enumerate}
    \end{proposition}

    \textcolor{black}{Proposition~\ref{37} characterizes the players' strategies in terms of the unit costs $\ba$ and $\bd$ as well as energy levels that influence $\dela$ and $\deld$. This result can be summarized in the ($\ba,\bd$) plane as shown in Fig. \ref{fig:fign2}. We will see later in a numerical example that the relation expressed in this plot holds for networks with more agents. In general, the player decides to attack (resp., to recover) if the unit cost $\ba$ (resp., $\bd$) is not too expensive. The attacker decides to attack for longer duration (Combined Strategy~3) if the attacker has large enough energy so that it is able to continue the attack for longer after the defender ends its recovery at $\taudf$.} 
    \vspace{-0.3cm}
    \subsection{\textcolor{black}{Discussion on the usage of $\widehat{\lambda}$}}
    {\color{black} In our formulation, the generalized edge connectivity $\widehat{\lambda}$ is used in the utilities of both players. This $\widehat{\lambda}$ captures the idea that some edges are weaker than others in connected graphs (and thus the attacker should attack the weakest edges while minimizing its energy usage). Moreover, some of the attacked edges are more crucial for the agents' communication than others (and thus the defender should recover the most important edges for the agents' communication). Among the different connectivity measures, the generalized edge connectivity is useful to characterize the resilience of the multi-agent systems represented by both connected and disconnected graphs.}
    
    \vspace{-0.1cm}
    \section{Application to Consensus Problem}
    In this section, a consensus problem of a multi-agent system \cite{ren,mesba,FB-LNS} in the face of jamming attacks is investigated. We apply our game approach to this problem.
   
    We assume that the graph $\mathcal{G}$ is connected and the agents communicate with neighbors continuously in time. Let $\mathcal{N}_i (t)$ be the set of neighbors of agent $i$, i.e., the agents sharing edges with agent $i$ at time $t$. Every agent $i$ has the scalar state $x_i$ whose dynamics are defined as
    \begin{equation} \label{state}
       \dot{x}_i(t) = \sum_{j\in \mathcal{N}_i(t)}(x_j(t)-x_i(t)), \quad x(0)=x_0, \quad t \geq 0,
    \end{equation}
    so that the state of all agents $x=[x_1~x_2\cdots x_n]^{\mathrm{T}}$ can converge to a consensus state $x_*$.
    
  We now introduce the notion of approximate consensus. Specifically,
for a given $\epsilon>0$, the approximate consensus set $\mathcal{D}_{\epsilon}\subset\mathbb{R}^{n}$
is given by $\mathcal{D}_{\epsilon} \coloneqq\{x\in\mathbb{R}^{n}\colon V(x)\leq\epsilon\},$
where 
\begin{align}
V(x) & \coloneqq\max_{i\in\mathcal{V}}x_{i}-\min_{i\in\mathcal{V}}x_{i},\quad x\in\mathbb{R}^{n}.\label{eq:Vdef}
\end{align}
 We characterize the effect of jamming attacks in terms of the time
for the agents to reach the approximate consensus set $\mathcal{D}_{\epsilon}$.
In particular, for the initial state $x(0)=x_{0}\in\mathbb{R}^{n}$,
the \emph{approximate consensus time} $T_{*}(x_{0})$ is given by
\begin{align}
T_{*} (x_{0})\coloneqq\inf\{t\geq0\colon x(t)\in\mathcal{D}_{\epsilon}\}.
\end{align}
 In our analysis, we also use the Laplacian matrix $L\in\mathbb{R}^{n\times n}$
associated with graph $\mathcal{G}$. Moreover, let $P\coloneqq e^{-\gamma^{\mathrm{A}}L}$
and $\underline{p} \coloneqq\max_{j\in\{1,\ldots,n\}}\min_{i\in\{1,\ldots,n\}}P_{i,j},\label{eq:underlinep}$, where $P_{i,j}$ denotes the $(i,j)$th entry of the matrix $P$. Notice that since $\mathcal{G}$ is connected and $\gamma^{\mathrm{A}}>0$, we have $P_{i,j}\in(0,1)$, and hence, $\underline{p}\in(0,1)$.

The next proposition gives an upper bound for the approximate consensus time of agents under jamming attacks. Here, we define $\lceil x \rceil$ as the ceiling function of $x$.
\vspace{-0.1cm}
\begin{proposition} \label{Proposition-Consensus} Consider the multi-agent
system (\ref{state}) with the initial condition $x_{0}\in\mathbb{R}^{n} \setminus D_\epsilon$.
Under the optimal attack and defense strategies for the resilient
graph game in Section III, the approximate consensus time
satisfies
\begin{align}
T_{*}(x_{0}) & \leq\frac{\beta^{\mathrm{A}}(\gamma^{\mathrm{A}}+\gamma^{\mathrm{D}})\left\lceil \frac{\ln\epsilon-\ln V(x_{0})}{\ln(1-\underline{p})}\right\rceil +\kappa^{\mathrm{A}}}{\beta^{\mathrm{A}}-\rho^{\mathrm{A}}}.\label{eq:proposition-result}
\end{align}
\end{proposition}
\vspace{-0.1cm}

Proposition~\ref{Proposition-Consensus} provides an upper bound related directly to the scalars $\beta^{\mathrm{A}}$,
$\kappa^{\mathrm{A}}$, $\rho^{\mathrm{A}}$ that characterize the
attacker's energy constraint, and the scalars $\gamma^{\mathrm{A}}$
and $\gamma^{\mathrm{D}}$ that respectively represent the attacker's
and the defender's waiting durations before taking actions in each
game. It is interesting to note that the attacker's energy parameters influence the bound more than the defender's energy parameters. In scenarios where there is no jamming attack (and hence no defense), from (\ref{eq:proposition-result}),
an upper bound of the approximate consensus time can be obtained as
$T_{*}(x_{0}) \leq (\gamma^{\mathrm{A}}+\gad)\left\lceil \frac{\ln\epsilon-\ln V(x_{0})}{\ln(1-\underline{p})}\right\rceil. \Big.$

\begin{figure}[t]
    %\hspace{-0.5cm}
    \centering
    \includegraphics[scale=0.55]{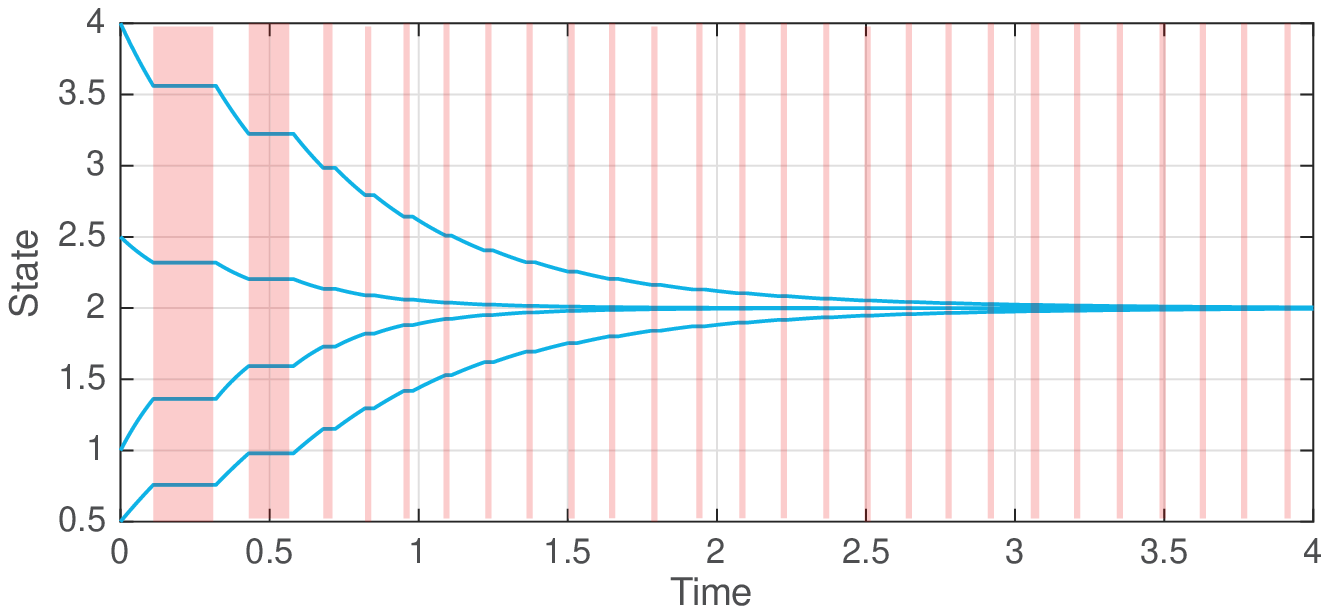}
    \vskip -5pt
    \caption{State trajectories with $\ka=0.5$ and $\ra=0.3$. \textcolor{black}{The red areas indicate the intervals where the attacker attacks.}}
    \label{fig:sim}
    \centering
    \includegraphics[scale=0.55]{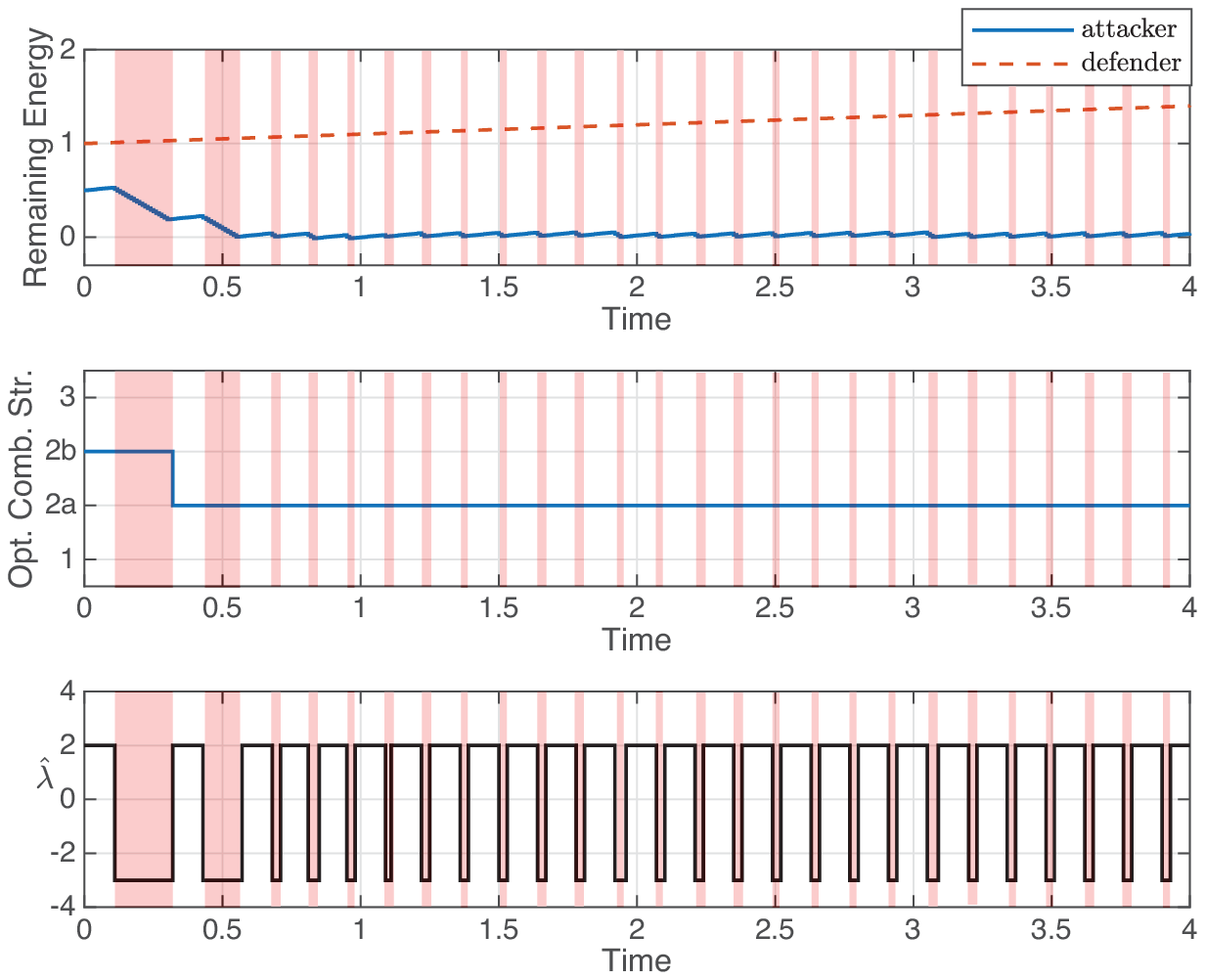}
    \vskip -15pt
    \caption{Remaining energy and optimal combined strategy for the two players, and the resulting $\widehat{\lambda}$ with $\ka=0.5$ and $\ra=0.3$. Note that the defender does not recover any edge, and hence the available energy for the defender accumulates continuously.}
    \label{fig:sim2}
    \vskip -15pt
\end{figure}
The approximate consensus time bound above
for the attack-free case is clearly smaller than in (\ref{eq:proposition-result})
when the attacker has positive energy resources ($\kappa^{\mathrm{A}},\rho^{\mathrm{A}}>0$)
and the defender has a nonzero initial waiting duration ($\gamma^{\mathrm{D}}>0$).
Note that with larger values of $\kappa^{\mathrm{A}}$ and $\rho^{\mathrm{A}}$,
the bound (\ref{eq:proposition-result}) becomes even larger, indicating
the possibility of slower consensus due to more damaging attacks. 
\vspace{-0.1cm}
\section{Numerical Examples}
In this section, we demonstrate the efficacy of the approach in the approximate consensus problem through numerical examples. 

We first compare the actual approximate consensus time for different energy parameters. We use the graph shown in Fig. \ref{fig:a} with $n=4$, and parameters $\ba=0.4$, $\bd=0.6$, $\kd=1$, $\rd=0.1$, $\gaa=0.1$, and $\gad=0.3$.

\begin{figure}
    %\hspace{-0.5cm}
    \vskip -15pt
    \centering
    \includegraphics[scale=0.55]{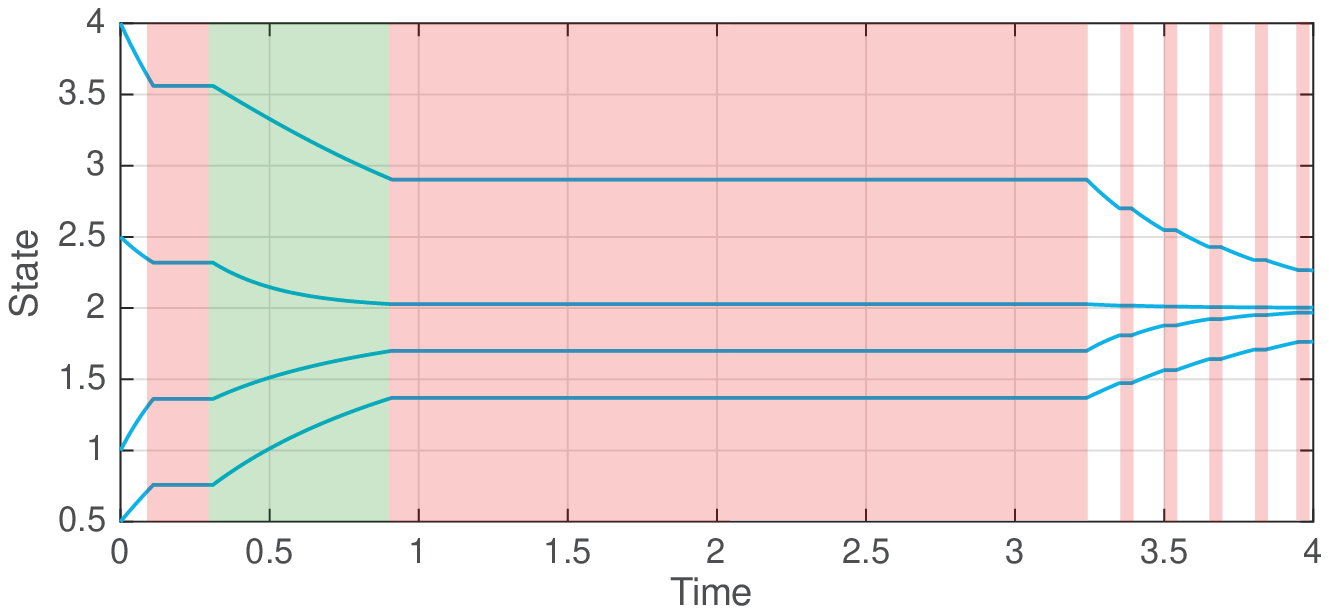}
    \vskip -5pt
    \caption{State trajectories with $\ka=5$ and $\ra=0.39$. \textcolor{black}{The green areas indicate the intervals where the defender recovers.}}
    \label{fig:sim3}
    \centering
    \includegraphics[scale=0.55]{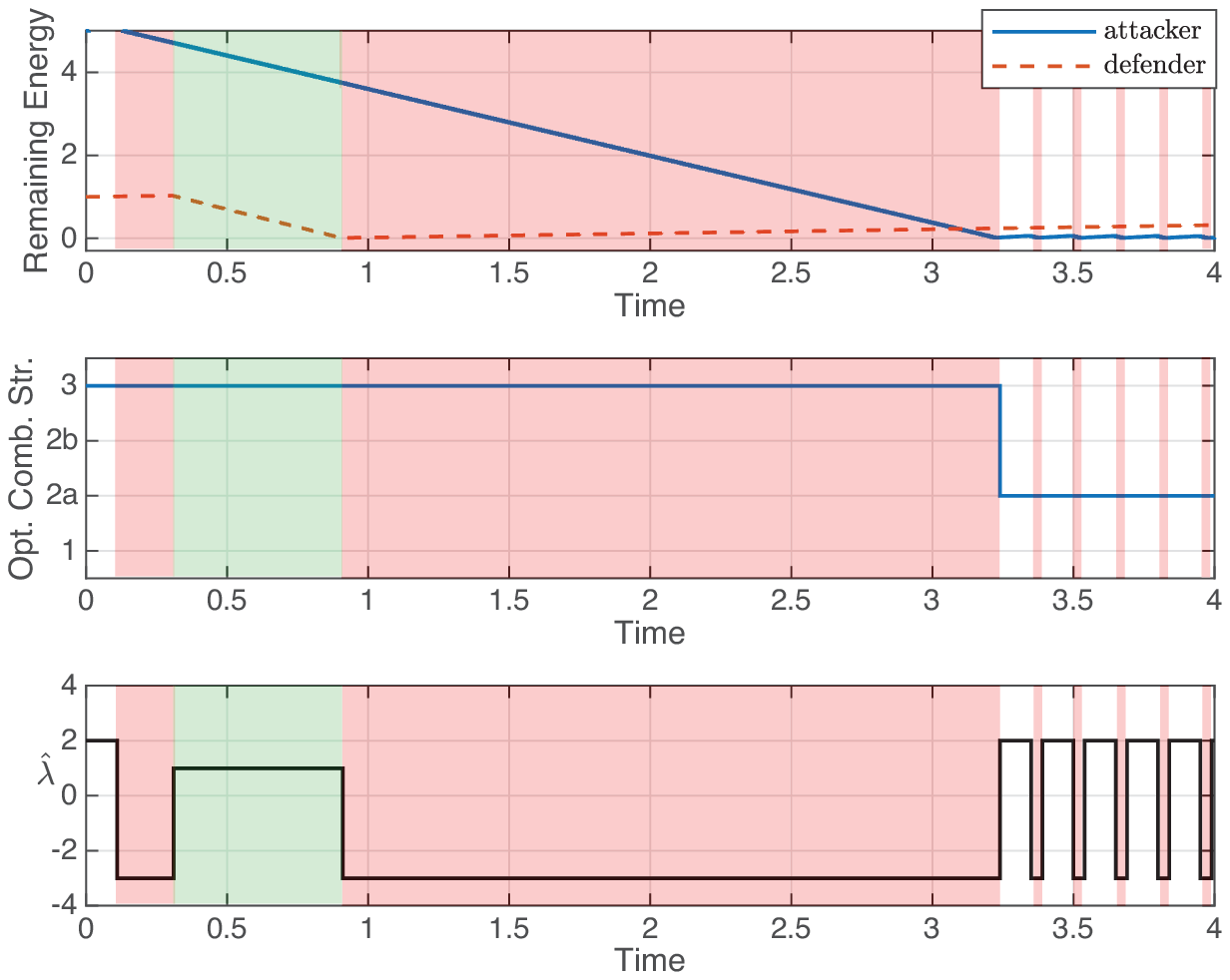}
    \vskip -15pt
    \caption{Remaining energy and optimal combined strategy for the two players, and the resulting $\widehat{\lambda}$. In this case, the attacker attacks all edges to achieve $\widehat{\lambda}(\mathcal{G}^{\mathrm{A}}_k)=-3$, where the defender recovers briefly in the first game to make the graph connected again.}
    \label{fig:sim4}
    \vskip -15pt
\end{figure}

    First, we use the parameters $\ka=0.5$ and $\ra=0.3$. Figs. $\ref{fig:sim}$ and $\ref{fig:sim2}$ show the states of the agents and properties of the players of the first simulation, with the agents eventually achieving approximate consensus at $t \approx 1.54$ with $\epsilon=0.5$. For comparison, when there is no jamming, it takes $t \approx 1.04$ to achieve the same level of approximate consensus. In the second simulation, we use the parameters $\ka=5$ and $\ra=0.39$. We present the results of this simulation in Figs. $\ref{fig:sim3}$ and $\ref{fig:sim4}$. It takes $t \approx 4$ with $\epsilon=0.5$ to achieve approximate consensus, which is longer than the first simulation because the attacker is given more energy. In these examples, the attacker decides to attack all edges, since by attacking more edges the defender has to recover more to increase the connectivity of the graph, which makes the recovery interval shorter.

\begin{figure}[t]
    \vskip -15pt
    \centering
    \includegraphics[scale=0.55]{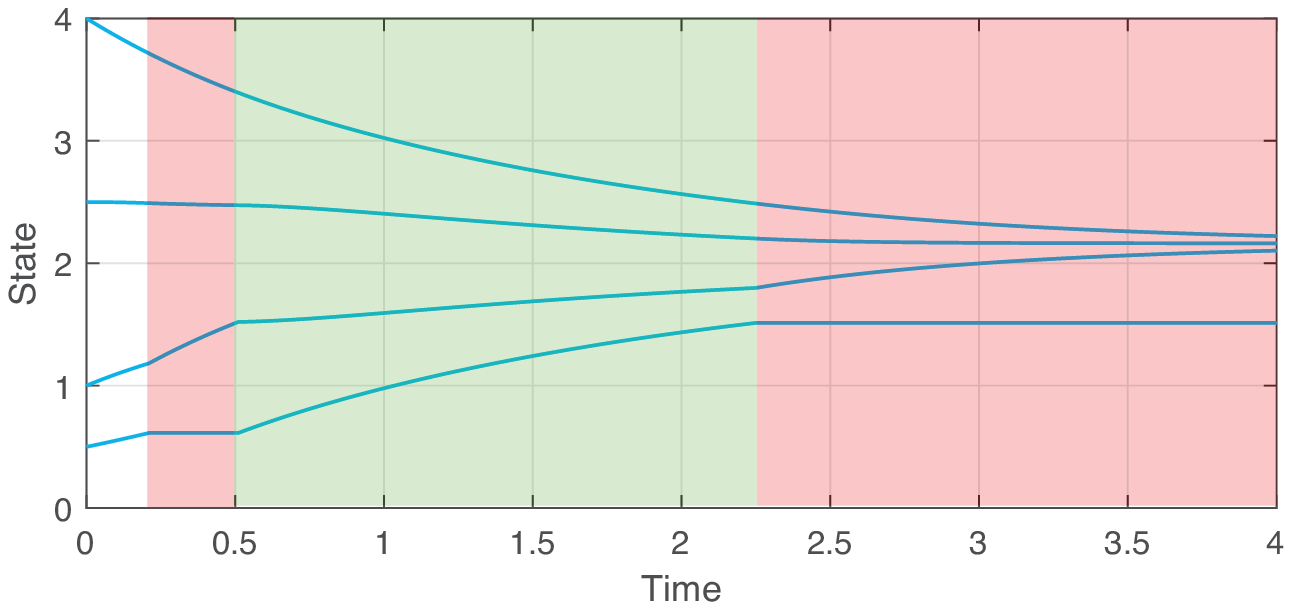}
    \vskip -5pt
    \caption{\textcolor{black}{State trajectories in the system with the path graph $\mathcal{G}$.}}
    \label{fig:sim5}
    \centering
    \includegraphics[scale=0.55]{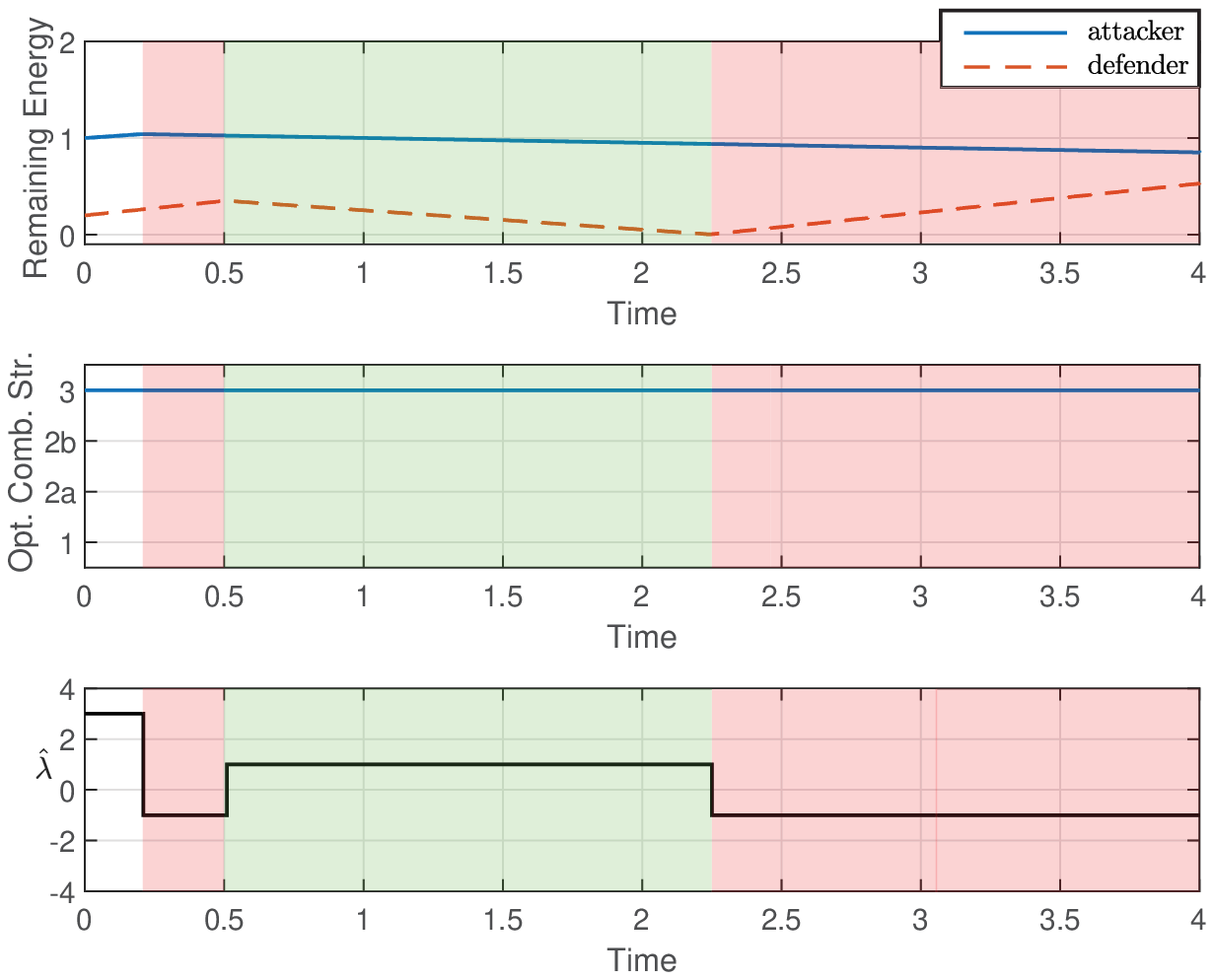}
    \vskip -15pt
    \caption{\textcolor{black}{Remaining energy and optimal combined strategy for the two players, and the resulting generalized edge connectivity in the system with the path graph $\mathcal{G}$.}}
    \label{fig:sim6}
\end{figure}

\begin{figure}[t]
    \centering
    \includegraphics[scale=0.55]{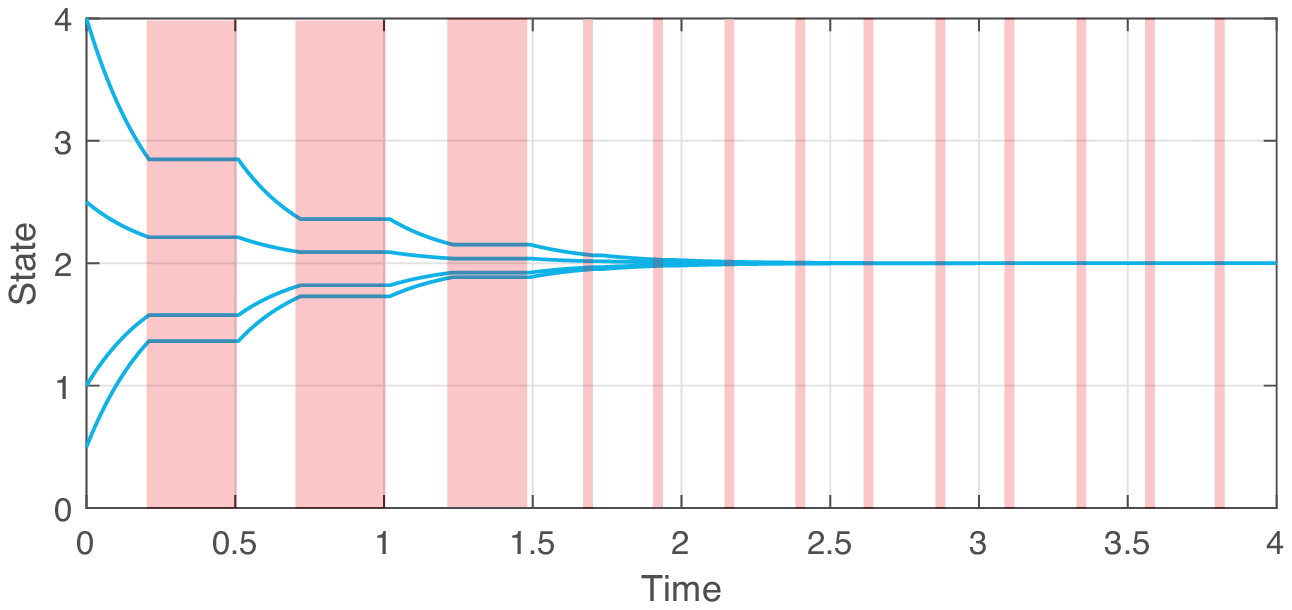}
    \vskip -5pt
    \caption{\textcolor{black}{State trajectories in the system with the complete graph $\mathcal{G}$.}}
    \label{fig:sim7}
    \centering
    \includegraphics[scale=0.55]{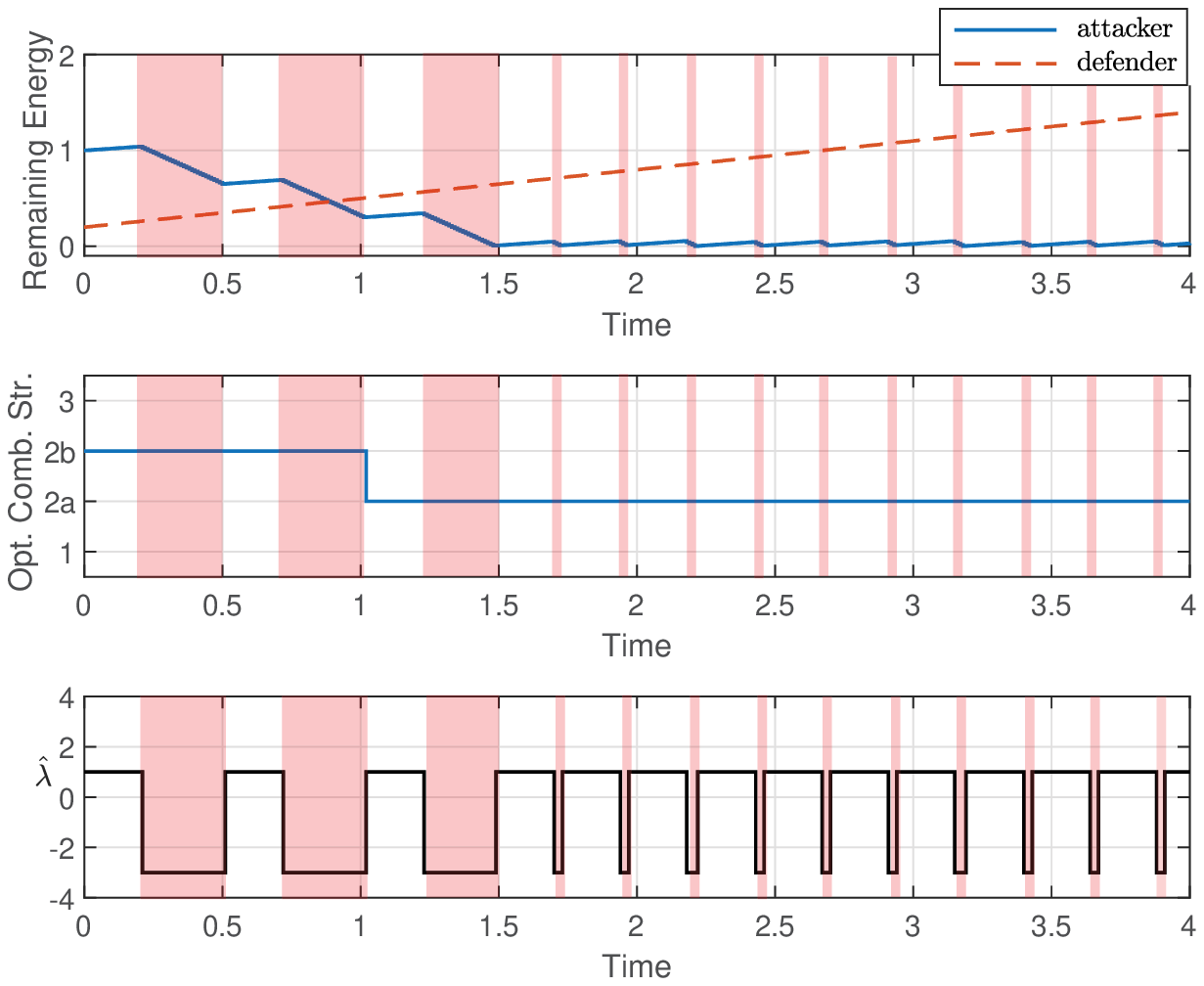}
    \vskip -15pt
    \caption{\textcolor{black}{Remaining energy and optimal combined strategy for the two players, and the resulting generalized edge connectivity in the system with the complete graph $\mathcal{G}$.}}
    \label{fig:sim8}
    \vskip -10pt
\end{figure}

    \textcolor{black}{Next, we compare the strategies of the players under different graph structures. Specifically, we run simulations on the path graph and the complete graph consisting of four nodes, while all other parameters are set to be the same across these two simulations. Fig. $\ref{fig:sim5}$ shows the state trajectory and Figs. $\ref{fig:sim6}$ shows the remaining energy, the optimal combined strategy, and the generalized edge connectivity versus time in the path graph. The corresponding results for the complete graph are shown in Figs. $\ref{fig:sim7}$ and $\ref{fig:sim8}$. We note that for the complete graph the attacker chooses to attack for shorter duration (Combined~Strategy~2b) due to the high connectivity of the graph structure. Specifically, the attacker needs to attack more edges (and hence takes more energy) to make the graph disconnected, and therefore the maximum attack interval becomes shorter compared to the attacks in the path graph. This shorter maximum attack duration results in a situation where the attacks on $[\taudf_k,\tauaf_k]$ interval are not able to compensate the negative payoff that the attacker receives on the $[\taudf_k,\tauaf_k]$ interval, causing the attacker to attack for only $\gad$ duration instead. Consequently, consensus is achieved faster in the complete graph than in the path graph. We can infer that graph structures influence the attack and recovery actions of the players, and graphs that have higher generalized edge connectivity are more resilient to attacks.}


\begin{figure}[ht]
    %\hspace{120pt}
    \centering
    \vskip 5pt
    \psfrag{A1a}{\small $\dela_k(1)$}
    \psfrag{A2a}{\small $\deld_k(1)$}
    \psfrag{A3a}{\scriptsize ($\ed_k=1$)}
    \psfrag{A4a}{\scriptsize ($\ed_k=3$)}
    \includegraphics[scale=0.7]{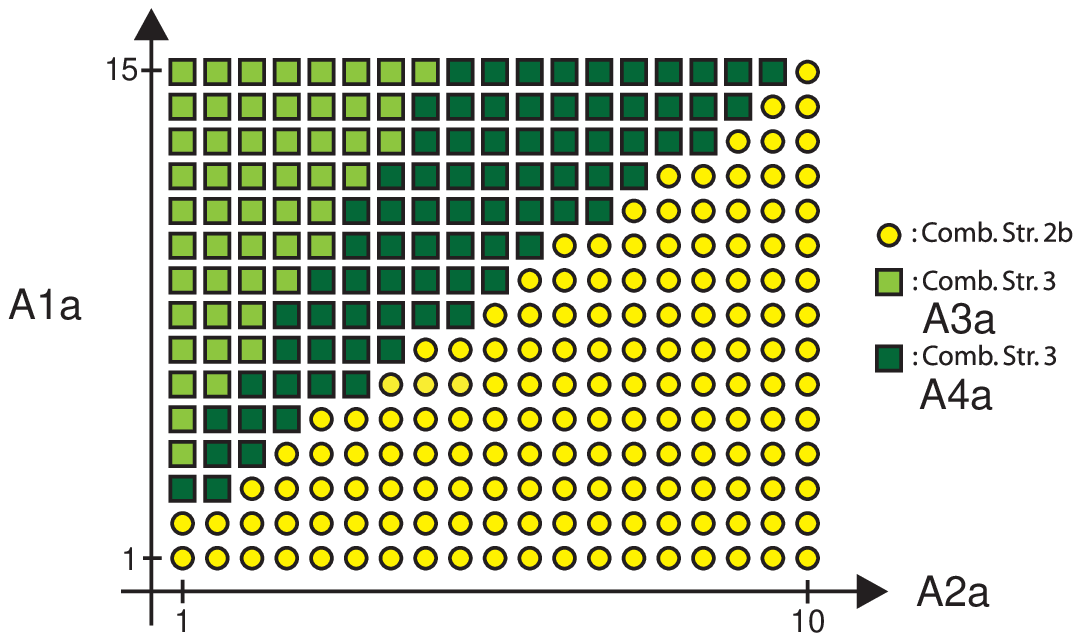}
    \vskip -5pt
    \caption{\textcolor{black}{Optimal combined strategies for different $\dela_k(\ea_k=1)$ and $\deld_k(\ed_k=1)$.}}
    \label{fig:sim9}
    \centering
    \vskip 5pt
    \psfrag{A1a}{\small $\ba$}
    \psfrag{A2a}{\small $\bd$}
    \includegraphics[scale=0.7]{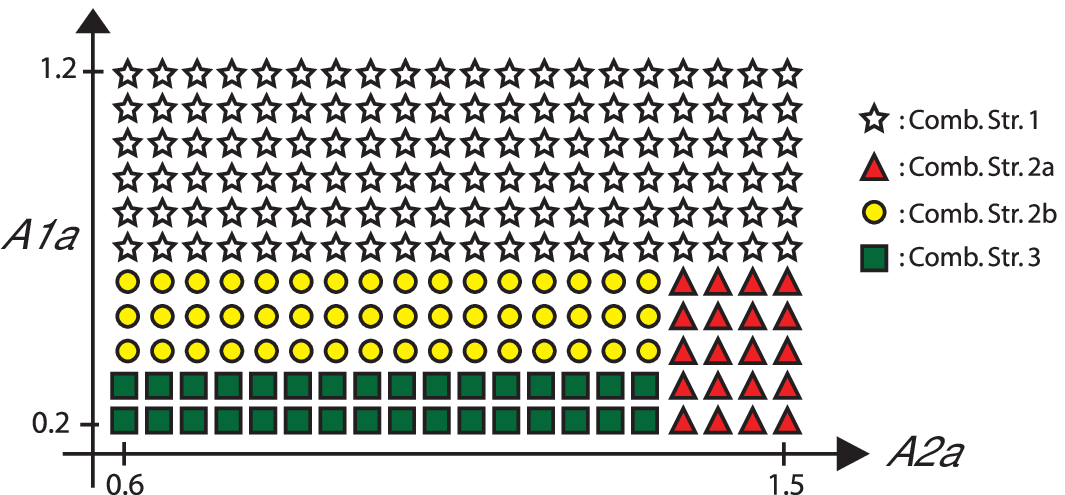}
    \vskip -5pt
    \caption{\textcolor{black}{Optimal combined strategies for different $\ba$ and $\bd$. The optimal numbers of edges are $\ea_k=5$, $\ed_k=3$ if the players decide to attack or recover.}}
    \label{fig:sim10}
    \vskip -15pt
\end{figure}

{\color{black} We also provide an example of how the energy, which affects the maximum attack/recovery durations, influences the equilibrium. We consider the graph in Fig. \ref{fig:a} with selected values of $\dela_k(\ea_k)$ and $\deld_k(\ed_k)$ in (\ref{en.a}) by changing the total consumed energy up to game $(k-1)$ represented as $\sum_{l=1}^{k-1} \ba\eam\inam$ and $\sum_{l=1}^{k-1} \bd\edm\indm$. The result is shown in Fig. \ref{fig:sim9} for $\ea_k=\ed_k=1$. In the figure, the yellow circles indicate that Combined~Strategy~2b is optimal with attacking five edges for given $\dela_k$ and $\deld_k$, whereas the green squares indicate that Combined~Strategy~3 is optimal with attacking five edges. The optimal strategy for the defender is to recover one edge and three edges (to make the graph connected again, e.g., $\{e_{12},e_{13},e_{34}\}$) in the areas with light green and dark green squares, respectively. The attacker attacks for longer durations if it possesses high amount of energy relative to the defender's energy. On the other hand, the defender with more energy will attempt to make the graph connected by recovering more edges. Fig. \ref{fig:sim9} can also be useful to estimate the equilibrium based on the past actions and energy parameters.

The optimal combined strategies for varying $\ba$ and $\bd$ are shown in Fig.~\ref{fig:sim10}. We note that Fig. \ref{fig:sim10} is similar to Fig.~\ref{fig:fign2} in terms of characterizing the influence of the unit costs $\ba$ and $\bd$ to the equilibrium, where the players tend not to attack or recover if the costs become higher. However, the critical values of $\ba$ and $\bd$ separating the optimal combined strategies in this set of simulations are lower than those found in Corollaries~\ref{corb} and~\ref{cora}. These critical values of $\ba$ and $\bd$ in the simulations are affected by generalized edge connectivity of $\mathcal{G}$ in Fig. \ref{fig:a}.}

\vspace{-0.1cm}
\section{Conclusion}

In this paper, we have considered resilient network problem in the context of multi-agent systems, formulated as a two-player game between the attacker and defender. Their utilities are determined by the communication among the agents. We fully characterized the optimal strategies of the players in terms of the edges and durations of action intervals. Several cases are possible to happen depending on the available energy of the players. For the consensus problem, we have shown that the time for the agents to reach approximate consensus will be delayed due to attacks by deriving an upper bound. 

\textcolor{black}{Note that in this paper, we have considered the generalized edge connectivity as one specific way to measure the network connectivity. In our recent paper \cite{yur2}, we consider clustering of agents in the network and also take account of the cluster sizes. It is also worth investigating other connectivity notions and non-uniform unit costs for practical applications. In \cite{yur3}, we have considered a problem formulation where not only the available energy but also the agents' states affect the results of the optimal strategies; this provides a more direct relation between the game and agents' dynamics.}


\appendices
\renewcommand{\thesectiondis}[2]{\Alph{section}:}
\section{Proof of Lemma~\ref{30}}
    The utility function in (\ref{ua}) can be rewritten as
   $\ua((\ea,\ina),(\ed,\ind)) = -\ha\ina-(\hd-\ha)\ind -\ba\ea\ina$. If there is recovery, i.e., $\ed,\ind > 0$, then $\hd>\ha$ according to the optimal strategy candidates. This implies that $-(\hd-\ha)\ind<-\ha\ind$ holds.
\vspace{-0.25cm}
\section{Proof of Lemma~\ref{31}}
    Substitute $\eaoc$ of (\ref{ex3}) into (\ref{hoa3}) to obtain
    $\faco = (\haoc-\hdoc)\xi +\fab(\eaoc)$. Since $\hdoc>\haoc$, it follows that $\faco \leq \fab(\eaoc)$, and therefore $\faco \leq \fabo$.
\vspace{-0.25cm}

\section{Proof of Lemma~\ref{32}}
    {\color{black}First, we show that $\text{sgn}(\fabo)=\text{sgn}(\fabbo)$. We can state $\fab(\ea)$ as $\fab(\ea) =\fabb(\ea) +(-\ha-\ba\ea)(\dela+\taua-\taud)$. By (\ref{aa}), we have $\taud \leq \tauaf$. Consequently, since $\tauaf \leq \taua+\dela$, we have $\dela+\taua \geq \taud$ for any possible $\dela$. Therefore, if $-\ha-\ba\ea>0$ is satisfied, then $\fab(\ea)>0$ and $\fabb(\ea)>0$, and vice versa. Again, since $\dela+\taua \geq \taud > \taua$, it follows that $\fabbo>0$ if and only if $\fabo>0$, since the attacker can always choose edges to make $\fabo$ and $\fabbo$ positive.} By a similar argument, $\fabbo<0$ if and only if $\fabo<0$. Thus, $\text{sgn}(\fabo)=\text{sgn}(\fabbo)$.
    
    {\color{black} Now, since $(-\haob-\ba\eaob)>0$ and $\dela+\taua \geq \taud$, it then follows that $\fabo \geq \fabbo$ if $\fabo>0$.}
\vspace{-0.25cm}
\section{Proof of Theorem~\ref{33}}
        We prove this result using the backward induction method. {\color{black} In Combined Strategy~1, recall that the attacker does not attack and the defender does not recover, so $\faa=\fda=0$. Therefore, the attacker chooses the optimal $\ea > 0$ to achieve positive utility.
        If the attacker attacks $\eao$, then the optimal strategy for the defender is to recover if and only if $\ud((\eao,\inao),(\edo,\indo > 0)) > \hao\dela$.}
        
        Recall that the utility of a player also depends on the other player's strategy. {\color{black} For example, if the defender's optimal strategy is to recover ($\ed > 0$) for given $\ea$, then the attacker's utility for given $\ea$ is $\ua((\ea,\ina),(\ed,\ind>0))$.}
        
        {\color{black} By backward induction, the six facts (i)--(vi) below hold:
        
        (i) From Lemmas \ref{31} and \ref{32}, since $\fabo > \faco$ and $\fabo$ has the same sign with $\fabbo$, Combined Strategy~1 is optimal if $\fabo<0=\hat{U}^{\mathrm{A}1}$, regardless of the defender's utility.} This fact proves point 1) in the theorem. 
        
        {\color{black} Since the case where $\fabo < 0$ is covered,  it is assumed that $\fabo \geq 0$ holds in all subsequent analysis for (ii)--(vi).
        
        (ii) Combined Strategy~2a with attacking $\eaob$ is the optimal combined strategy if $\fdcdua=\ud((\eaob,\dela),(\edo,\deld>0))$ is less than $\fdb=\ud((\eaob,\dela),(0,0))$, since the defender chooses not to recover ($\edo=0$) and $\fabo=\ua((\eaob,\dela),(0,0))$ is the maximum possible utility for the attacker from Lemmas \ref{31} and \ref{32}.} This fact corresponds to point 2)a) in the theorem. 
        
        {\color{black} Since the case where $\fdcdua<\fdb$ is covered, beginning from (iii) to (vi), it is further assumed that $\fdcdua \geq \fdb$, i.e., the defender chooses to recover from $\eaob$. Since $\edob > 0$ and $\fabo=\ua((\eaob,\dela),(0,0))$, in the subsequent cases, the attacker's optimal number of edges are not $\eaob$ (which corresponds to $\fabo$).
        In (iii) and (iv), we analyze the case where $\faco\geq \fabbo$, which means that Strategy~A3 yields more or equal utility than Strategy~A2b for the attacker.}
        
        (iii) Due to the possible jump between $\hd=-1$ to $\hd=1$ by recovering only one edge, the defender may have different optimal strategies (whether to recover or not) given different attacked edges. {\color{black} From Lemma \ref{30}, since the attacker has better utility if the defender does not recover, here the attacker's optimal strategy is to attack $\eato$ if $\facdua=\ua((\eato,\dela),(0,0)) $ is greater than $\faco = \ua((\eaoc,\dela),(\ed,\deld>0))$, with $\eato$ being the optimal number of edges among the edges that cannot be recovered if attacked, as in (\ref{38})}. Therefore, Strategies~A1, A2b, and A3 are not optimal. This corresponds to point 2)b)II) in the theorem.
        
        {\color{black} (iv)  Otherwise, Combined Strategy~3 (point 4) in the theorem) is the optimal combined strategy if $\facdua\leq \faco$. Here, the defender's optimal strategy is to recover if the attacker attacks $\eaoc$. Since $\faco \geq \max\{\fabbo,0\}$, the attacker has better utility than in Strategies~A1, A2a, and A2b. 
        
        In (v) and (vi), we analyze the case where $\fabbo > \faco$.
        
        (v) Similar as in (iii), Combined Strategy~2a is the optimal strategy if $\facdua \geq \fabbo$.} In this case, the attacker has better utility than in Strategies~A1,~A2b, and A3. {\color{black} However, since $\fdcdua \geq \fdb$, the attacker does not attack $\eaob$. This fact corresponds to point 2)b)I) in the theorem.
        
        (vi) If $\facdua < \fabbo$, Strategy A2b is the optimal strategy for the attacker since $\fabbo > \max(\faco,\facdua)$ and utility $\fabo$ cannot be achieved because $\fdcdua\geq \fdb$.} This corresponds to point 3) in the theorem.
\vspace{-0.25cm}
\section{Proof of Corollary~\ref{corb}}
      {\color{black} In Strategy~D3, since $\min\{\deld,\dela+\taua-\taud\} > 0$, the necessary condition for Strategy~D3 to be the optimal strategy is $\bd<(\hdoc-\haoc)/\edoc$, i.e., the cost of recovering edges is not too large. If this condition is not satisfied, then it is better for the defender not to recover as in Strategy~D1. By recovering one edge the defender is able to make $(\hd-\ha)/\ed=2$ at most. Thus, if $\bd > 2$, then the defender does not recover any edge.}
\vspace{-0.25cm}
\section{Proof of Corollary~\ref{cora}}
     {\color{black} From Theorem \ref{33}, the attacker decides to attack if $\fab(\eaob) \geq 0$. Since $\dela > 0$, Strategy~A2a is the optimal strategy if $-\haob-\ba\eaob\geq 0$, assuming that the defender cannot recover. By Lemma~3.1, $\fab(\eaob)>\fac(\eaoc)$, and thus Strategy~A1 is the optimal strategy if $-\haob-\ba\eaob<0$. 
    
    Since $\ba\eaob > 0$, to make $-\haob-\ba\eaob>0$, it must hold that $\haob<0$. Therefore, the attacker must attack enough edges to make $\ga$ disconnected.
    Because $-\haob/\eaob$ cannot exceed 1, in order to obtain positive utility, $\ba \leq 1$ must be satisfied.}
\vspace{-0.25cm}
\section{Proof of Proposition~\ref{37}}
        {\color{black} Since $|\mathcal{E}|=1$, the following four facts corresponding to points 1) to 4) in Theorem \ref{33} hold:
        
        (i) Combined Strategy~1 is optimal if $\fabo<0$. Since $|\mathcal{E}| \ =1$, $\ha=-1$ is always true if $\ea>0$. From (\ref{hoa}), it is clear that $\fabo<0$ if $\ba>1$.
        
        (ii) In order for Combined Strategy 2a to be optimal, a common condition is that $\fabo\geq0$, which holds if $\ba\leq1$. The condition $\fdcdua<\fdb$ then holds if $\bd>2$. Note that $\mathcal{M}$ consists of $|\mathcal{E}|=1$ if $\bd> 2$ and empty otherwise. Hence, $\facdua=\fabo$ holds if $\bd>2$, otherwise $\facdua=0$ holds. Therefore, in point 2)b) in Theorem~\ref{33}, condition $\fdcdua\geq\fdb$ implies that $\facdua=0$ holds, which means that the conditions 2)b)I) and 2)b)II) cannot be satisfied (from Lemma \ref{32}). 
        
        (iii) Combined Strategy~2b is optimal if $\fabo\geq 0$, which holds if $\ba\leq1$. The other condition is that $\fdcdua\geq\fdb$, which holds if $\bd\leq2$. Conditions $\fabbo>\facdua$ is always true (see point (ii) in this proof above). With $n=2$, condition $\fabbo>\faco$ is true if $\ba > 1-\frac{2\xi}{\dela-\taud+\taua}$ holds, with $\xi$ defined in (\ref{xi}).
        
        (iv)  It then follows that Combined Strategy~3 is optimal if $\fabo\geq 0$ (holds if $\ba\leq1$), $\fdcdua \geq \fdb$ (holds if $\bd\leq2$), and $\fabbo\leq\faco$ (holds if $\ba \leq 1-\frac{2\xi}{\dela-\taud+\taua}$), under which the condition $\faco\geq\facdua$ holds.}

\vspace{-0.25cm}
\section{Proof of Proposition~\ref{Proposition-Consensus}}
\vspace{-0.0cm}
The agents do not face any attacks during the intervals $[\underline{t}_{k},\underline{\tau}_{k}^{\mathrm{A}})$,
$k\in\mathbb{N}$. Thus, from (\ref{state}), $\dot{x}(t) =-Lx(t),\ t\in[\underline{t}_{k},\underline{\tau}_{k}^{\mathrm{A}}),\ k\in\mathbb{N}$. Noting that $\underline{\tau}_{k}^{\mathrm{A}}=\underline{t}_{k}+\gamma^{\mathrm{A}}$,
we obtain $x(\underline{\tau}_{k}^{\mathrm{A}}) =Px(\underline{t}_{k}),\ k\in\mathbb{N}.$
Now by using Lemma~12.8 of \cite{FB-LNS}, it follows that 
\begin{align}
V(x(\underline{\tau}_{k}^{\mathrm{A}})) & =V(Px(\underline{t}_{k}))\leq(1-\underline{p})V(x(\underline{t}_{k})).\,\,\label{eq:V-relation-1}
\end{align}

During the intervals $[\underline{\tau}_{k}^{\mathrm{A}},\underline{t}_{k+1})$,
$k\in\mathbb{N}$, there may be attacks and the communication between
certain agents may be jammed. \textcolor{black}{It then follows from (\ref{state}) that}
\begin{align}
V(x(\underline{t}_{k+1})) & \leq V(x(\underline{\tau}_{k}^{\mathrm{A}})),\quad k\in\mathbb{N}.\label{eq:V-relation-2}
\end{align}
By (\ref{eq:V-relation-1}) and (\ref{eq:V-relation-2}), $V(x(\underline{t}_{k+1}))\leq(1-\underline{p})V(x(\underline{t}_{k}))$,
and thus, 
\begin{align}
V(x(\underline{t}_{k+1})) & \leq(1-\underline{p})^{k}V(x(\underline{t}_{1}))=(1-\underline{p})^{k}V(x_{0}),\, k\in\mathbb{N} .\label{eq:V-k-relation}
\end{align}
Let $k_{*}\coloneqq\left\lceil(\ln\epsilon-\ln V(x_{0}))/\ln(1-\underline{p})\right\rceil$.
By (\ref{eq:V-k-relation}), it clearly holds
$V(x(\underline{t}_{k_{*}+1}))\leq\epsilon$, and therefore,
\begin{align}
x(t) & \in\mathcal{D}_{\epsilon},\quad t\geq\underline{t}_{k_{*}+1}.\label{eq:tkstarplusone}
\end{align}

Our next goal is to find an upper bound of $\underline{t}_{k_{*}+1}$.
First, by the energy constraint for the attacker given in (\ref{a}), $\beta^{\mathrm{A}}\sum_{k=1}^{k_{*}}\ea_k\delta_{k}^{\mathrm{A}} \leq\kappa^{\mathrm{A}}+\rho^{\mathrm{A}}\underline{t}_{k_{*}+1}$ holds.
As indicated by the optimal strategies derived in Theorem~\ref{33},
$\ea_k=0$ implies that $\delta_{k}^{\mathrm{A}}=0$.
Hence, we have $\ea_k\delta_{k}^{\mathrm{A}}\geq\delta_{k}^{\mathrm{A}}$, which implies
\begin{align}
\sum_{k=1}^{k_{*}}\delta_{k}^{\mathrm{A}} & \leq\frac{1}{\beta^{\mathrm{A}}}\beta^{\mathrm{A}}\sum_{k=1}^{k_{*}}\ea_k\delta_{k}^{\mathrm{A}}\leq\frac{\kappa^{\mathrm{A}}}{\beta^{\mathrm{A}}}+\frac{\rho^{\mathrm{A}}}{\beta^{\mathrm{A}}}\underline{t}_{k_{*}+1}.\label{eq:energy-result}
\end{align}
 Next, by (\ref{delt}), 
\begin{align}
\underline{t}_{k+1}=\overline{t}_{k} & \leq\underline{t}_{k}+\gamma^{\mathrm{A}}+\gamma^{\mathrm{D}}+\delta_{k}^{\mathrm{A}},\quad k\in\mathbb{N}.\label{eq:tkupperbound}
\end{align}
It then follows from (\ref{eq:energy-result}) and (\ref{eq:tkupperbound})
that $\underline{t}_{k_{*}+1} =\sum_{k=1}^{k_{*}}(\underline{t}_{k+1}-\underline{t}_{k}) \leq(\gamma^{\mathrm{A}}+\gamma^{\mathrm{D}})k_{*}+\frac{\kappa^{\mathrm{A}}}{\beta^{\mathrm{A}}}+\frac{\rho^{\mathrm{A}}}{\beta^{\mathrm{A}}}\underline{t}_{k_{*}+1},$
and hence, 
\begin{align}
\underline{t}_{k_{*}+1} & \leq\frac{(\gamma^{\mathrm{A}}+\gamma^{\mathrm{D}})\left\lceil \frac{\ln\epsilon-\ln V(x_{0})}{\ln(1-\underline{p})}\right\rceil +\frac{\kappa^{\mathrm{A}}}{\beta^{\mathrm{A}}}}{1-\frac{\rho^{\mathrm{A}}}{\beta^{\mathrm{A}}}}.\label{eq:tkstaroneupperbound}
\end{align}
Finally, by (\ref{eq:tkstarplusone}) and (\ref{eq:tkstaroneupperbound}),
we obtain (\ref{eq:proposition-result}).
% or






%\begin{thebibliography}{9}

\nocite{*}
\vspace{-0.1cm}
\bibliographystyle{IEEEtran}
\bibliography{IEEEabrv,tesv2}

\end{document}